\documentclass[graybox]{svmult}

\usepackage{type1cm}        % activate if the above 3 fonts are
\usepackage{makeidx}         % allows index generation
\usepackage{graphicx}        % standard LaTeX graphics tool
\usepackage{multicol}        % used for the two-column index
\usepackage[bottom]{footmisc}% places footnotes at page bottom

\usepackage{graphicx}

\usepackage{newtxtext}       % 
\usepackage{newtxmath}       % selects Times Roman as basic font

\makeindex             % used for the subject index
\begin{document}

\title*{The universality of self-organisation -- a path to an atom printer?}
\author{Serim Ilday, F. \"{O}mer Ilday}
\institute{Serim Ilday \at UNAM – National Nanotechnology Research Center and Institute of Materials Science and Nanotechnology, Bilkent University, Ankara, 06800, Turkey, \email{serim@bilkent.edu.tr}
\and F. \"{O}mer Ilday \at Physics Department, and Electrical and Electronics Engineering Department, Bilkent University, Ankara, 06800, Turkey \email{ilday@bilkent.edu.tr}}
\maketitle

%\abstract*{.}

\abstract{More than thirty years ago, Donald Eigler and Erhard Schweizer spelt the letters IBM by positioning 35 individual Xenon atoms at 4 Kelvin temperature using a scanning tunnelling microscope. The arrangement took approximately 22 hours. This was an outstanding demonstration of control over individual atoms. Since then, 3D printers developed into a near-ubiquitous technology. Nevertheless, with typical resolutions in the micrometres, they are far from the atomic scale of control that the IBM demonstration seemed to herald. Even the highest resolution achieved with ultrafast lasers driving two-photon polymerization barely reaches 100 nm, three orders of magnitude distant from the atomic scale. Here, we adopt a long-term view when we ask about the possibility of a 3D atom printer, which can build an arbitrarily shaped object of macroscopic dimensions with control over its atomic structure at room temperature and within a reasonable amount of time. After discussing the state-of-the-art technology based on direct laser writing, we identify three fundamental challenges to overcome. The first is the \emph{fat fingers problem}, which refers to laser wavelengths being much larger than the size of the atoms. The second one is \emph{complexity explosion}; namely, the number of processing steps scales with the inverse cube of the resolution, leading to prohibitively long processing times. The third challenge is the increasing strength of random fluctuations as the size of the smallest volume element to be printed approaches the atomic scale. This requires control over the fluctuations, which we call \emph{mischief of fluctuations}. Although direct-writing techniques offer sufficient resolution, speed, and excellent flexibility for the mesoscopic scale, each of the three fundamental problems above appears enough to render the atomic scale unreachable. In contrast, the three challenges of direct writing are not fundamental limitations to self-organisation. This chapter proposes a potential path to a 3D atom printer, where laser-driven self-organisation can complement direct-writing techniques by bridging the atomic and mesoscopic scales.}

\section{Introduction}
\label{sec:1}
\begin{figure}[t]
\includegraphics[width=11.6cm]{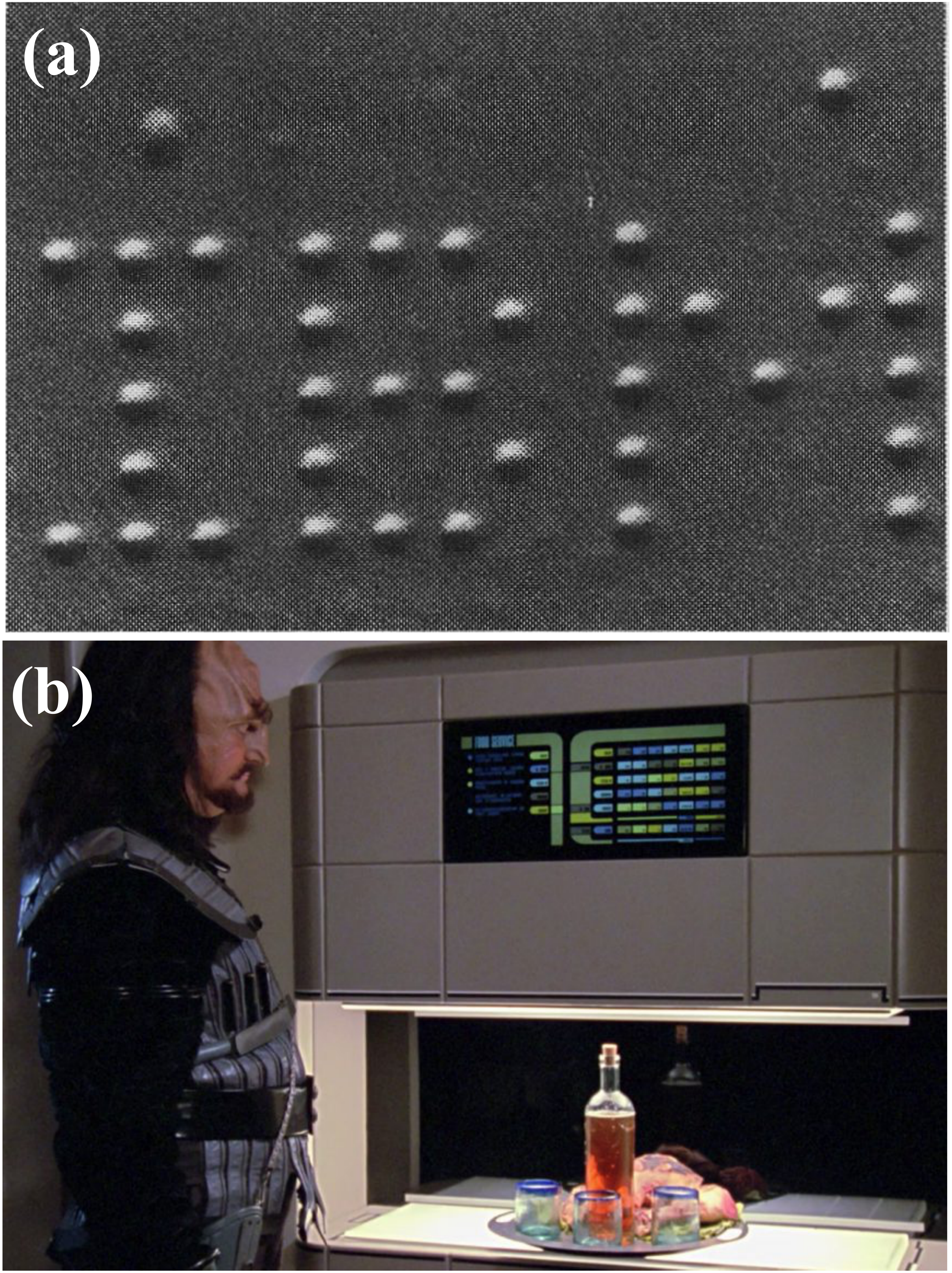}
\caption{The letters IBM were spelt with 35 Xe atoms in 1990 \cite{1} (Image Credit: IBM). (b) The image of a plate of food synthesized by a Replicator as imagined in the science-fictional production, Star Trek (Image Credit: Paramount Television).}
\label{fig:IBM_Star_Trek}       
\end{figure}

In a 1990 paper published in {\em Nature}, D. Eigler and E. Schweizer demonstrated the astounding capability of controlling individual atoms \cite{1}. Their demonstration of spelling the letters IBM with 35 Xe atoms was revolutionary, inspiring, and raised high hopes that we might, after all, assemble custom-designed structures and functionalities atom-by-atom (fig. \ref{fig:IBM_Star_Trek}a). Around the same time, this vision was popularised through a futuristic device called the Replicator in the science-fiction TV series {\em Star Trek: The Next Generation} (fig. \ref{fig:IBM_Star_Trek}b). A \emph{Replicator} is a fictional machine that can synthesise any inanimate object with molecular precision. In principle, the spelling of IBM with 35 Xe atoms constituted a solid scientific basis for anticipating a \emph{Replicator} being built in the future. Three decades after the IBM demonstration, so-called 3D printers that can create arbitrarily shaped objects form a mature technology, some of which are, incidentally, branded as \emph{Replicators}.

\begin{figure}[t]
%\sidecaption
% For example, with the graphicx style use
\includegraphics[width=\textwidth]{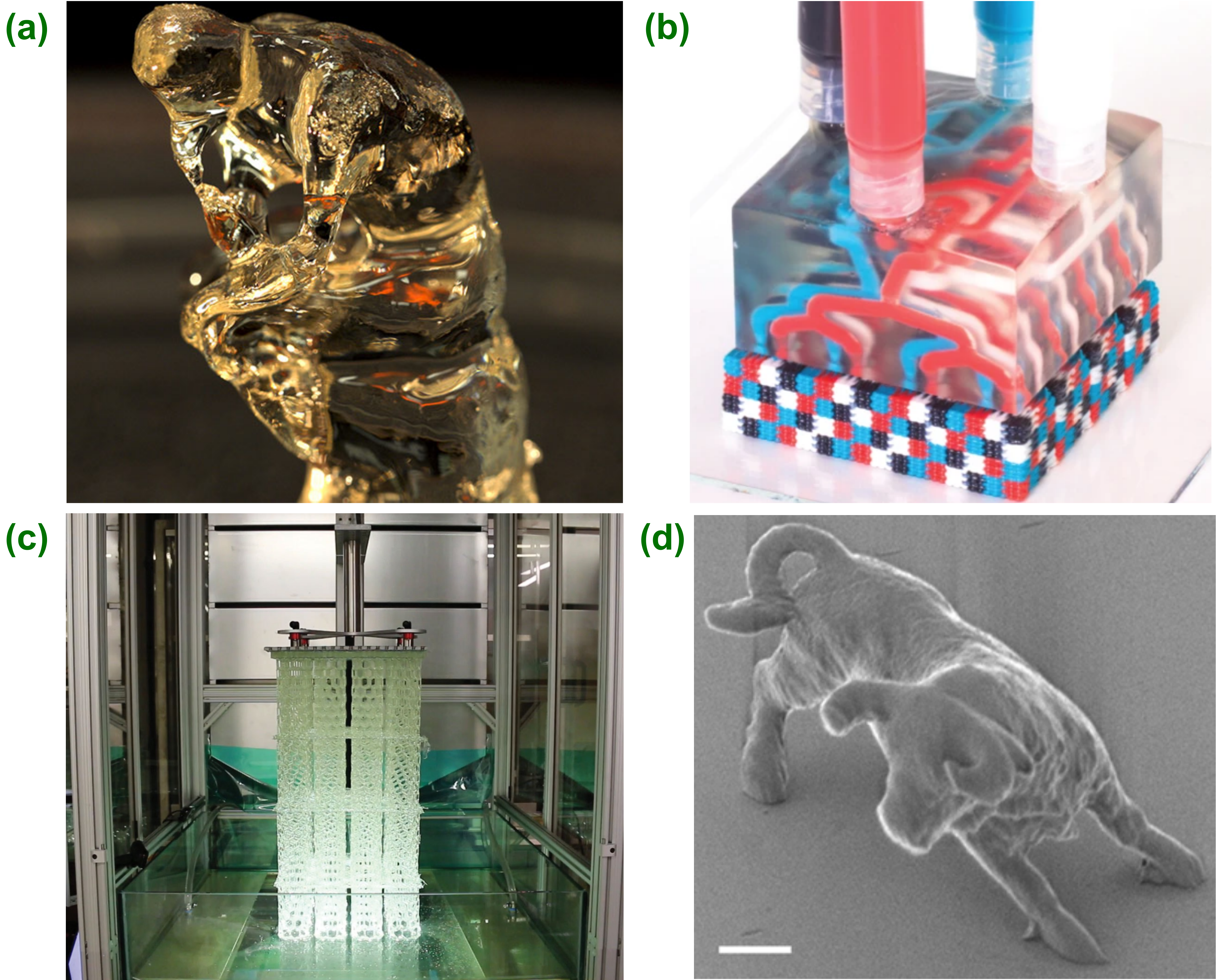}
%
% If no graphics program available, insert a blank space i.e. use
%\picplace{5cm}{2cm} % Give the correct figure height and width in cm
%
\caption{(a) 3D printed 40 millimetres-tall version of Auguste Rodin’s The Thinker using computed axial lithography \cite{13} (Image Credit: University of California at Berkeley). (b) Multilateral multi-nozzle 3D printing of several millimetres-large voxelated structures using four different materials \cite{14} (Image Credit: Harvard University). (c) Large-volume 3D printing of a meters-long polyurethane acrylate lattice from a liquid interface \cite{15} (Image Credit: Northwestern University). (d) Micro-nano fabrication of a bull structure at sub-diffraction-limit resolution using two-photon polymerization technique (Scale bar 2 $\mu$m) \cite{16}. Credit: Osaka University.}
\label{fig:3D}       % Give a unique label
\end{figure}

However, present-day 3D printers can print using typically single or a few different materials, mainly metal alloys and plastics. Their resolutions are in the 20 $\mu$m to 1 mm range. The better resolutions are obtained using laser-based melting of the material, limiting them mainly to metals and metal alloys despite ongoing efforts to broaden the material compatibility \cite{2,3,4,5,6,7}. Only the most sophisticated printers based on ultrafast lasers driving multiphoton-induced polymerisation can reach the mesoscopic scale with resolutions ranging from $\sim150$ nm \cite{8,9,10} to a micron \cite{11,12}. Unfortunately, the multiphoton polymerisation technique is even more limited in the materials it can print. Examples of the state-of-the-art 3D printing methodologies are provided in fig. \ref{fig:3D} \cite{13,14,15,16}, which were also discussed in a recent feature article \cite{17}.

The current state of the art is far from the possibilities heralded by the IBM demonstration or the fictional {\em Replicator} technology. They fail most strikingly in two aspects, spatial resolution and limited material choices. Despite the significant interest and rapid progress in 3D printing, it is an open question whether reaching the atomic scale and universality in materials is merely a matter of more time and better engineering.

Can we anticipate progress that eventually leads to the capability to print a complex structure, perhaps, a biological tissue, by assembling it atom-by-atom? Any biological tissue comprises enormous complexity, organised into different structures from the tissue level to the individual cells, which, in turn, contain sub-cellular structures that are themselves complex combinations of hundreds of biomolecules. These wildly varying structures exist in a state of homeostasis, locally resisting the deleterious effects of the second law of thermodynamics through continuous processing of matter and an energy flux \cite{18}. They are dynamic and adaptive, capable of autocatalysis, autoinhibition, self-healing, self-replication, regeneration, and reproduction. Unlike the atoms spelling out IBM, which could only be maintained in that configuration by cooling the sample to 4 K, biological tissue thrives at temperatures in the range of 300 K, where the thermal fluctuations are extreme at the atomic level \cite{19}.

Nevertheless, the molecular machinery that maintains a 1-mm$^3$ slab of tissue is well adapted to, and in fact, evolved to exploit these fluctuations \cite{20}. It is imperative to recognize that nothing about its structure or molecular machinery is controlled precisely down to the individual atom. Some $10^{21}$ atoms participate in this dynamic steady state, and their assembly and continuous rearrangements are far from error-free. Still, a diverse network of feedback loops called the metabolism keeps both the structure and the machinery running successfully \cite{21}. Even if we ignore these amazing dynamic features, its vast structural complexity exceeds, by a considerable margin, anything that we have been able to manufacture to date.

Our purpose in this Chapter is to provide a perspective for a possible pathway that could eventually enable a technology to print arbitrarily shaped objects of macroscopic dimensions with control over its atomic structure at room temperature, a 3D atom printer, if you will. We will argue that recent developments in laser physics and technology and self-organization methods can be exploited to merge top-down and bottom-up approaches into a single device that could bridge the gap between the atomic and mesoscopic scales. Section \ref{sec:2} will give a brief introduction to the state-of-the-art direct laser writing technology and bring forward the possibility of merging additive and subtractive manufacturing methods using, what we refer to as the \emph{Omni-modality} approach. Section \ref{sec:3} will stress that regardless of advanced engineering, the top-down manufacturing methods will always be limited in reaching the atomic scale by the three fundamental challenges and bring forward laser-driven self-organisation techniques to help overcome these challenges and complement them. Finally, Section \ref{sec:4} will give a perspective on an atom printer’s practicality, design, and potential and outline its working principles that combine top-down and bottom-up approaches.

\section{The State-of-the-Art Direct Laser Writing Technology}
\label{sec:2}

The fabrication of the majority of small-scale engineered structures is based on transferring or impressing a structure from a previously prepared mould or mask to the target material or using direct, point-by-point processing of the material(s) involved to define the structure. The former defers the problem to creating a mould or a mask. This approach works best (and exceptionally well for microelectronics) when many identical copies of a single structure are made. The latter can be further divided into additive or subtractive techniques. While many alternative technologies compete in both categories, arguably the best-known methods are conventional mechanical drilling (limited to $>\!\!100\;\mu$m scale) and laser removal for the subtractive approach \cite{22,23,24}. Most additive manufacturing solutions rely on low-cost, low-resolution 3D printers that melt and resolidify plastic derivatives for the consumer market and laser-based 3D printers that generally build out of metal alloys for the professional market \cite{25,26}. Both subtractive and additive manufacturing techniques work best in the micron scale and above, and the mesoscopic scale of $0.1-1 \mu$m continues to be the domain of high-end niche applications. In addition to the additive and subtractive techniques, a third possibility is what we refer to as the {\em Omni-modality} approach, where both methods are seamlessly interchanged within the same manufacturing platform. This approach is best suited to laser technology, where recent inventions \cite{27} make it practical to use a single laser for both processes. Despite the convenience of this approach, it alone does not remove the limitations on the resolution that additive and subtractive techniques suffer from, which will be discussed in Section \ref{sec:3}.

\subsection{Additive Manufacturing}
\label{subsec:2}

Additive manufacturing is regarded as an essential component of the so-called $4^{\rm th}$ Industrial Revolution, speeding time to market, enhancing productivity, flexibility, and reducing the use of resources. While it is a fast-growing area, the share of additive manufacturing in the global market remains low. This is due to various challenges, such as the high cost of the investment; lack of fast, scalable, serial production workflows; issues with repeatability, quality, and standardisation; and limited possibilities for materials to be printed. While many of these immediate limitations will likely be solved with further engineering effort, the restriction to printing resolution stands out as a fundamental problem. By printing resolution, we refer to the size of the smallest independently printable volume element or the voxel size. In additive manufacturing, which is based on direct writing, a lower limit to the voxel size is set by the size of the energy source activating the process, which is the spot size of the optical beam in the case of laser-based 3D printers. This, in turn, is limited by diffraction to a few 100 nm using optical wavelengths. In the vast majority of the cases, the practical resolution limits are much larger, at 10 $\mu$m or even larger. A continuous-wave (CW) laser is commonly used as a precise heat source to melt or sinter the target material, which may be in wire or powder form. Technical factors, such as the size of the powder particles or the diameter of the wire, set the printing resolution. With these techniques, currently achieved resolutions are $>\!\!20\!-\!30\;\mu$m for the powder method and $>\!\!200\;\mu$m for the wire method, which are much higher than the fundamental limits set by optical diffraction.

The highest resolutions in 3D printing are achieved with the so-called multiphoton polymerisation technique \cite{28}. This technique uses a polymer that is initially in liquid form. It can be solidified through a polymerisation process using light of sufficient photon energy, often near the ultraviolet (UV). For the activation of the polymerisation process, a laser with twice or even triple of the activation wavelength is used. This way, the activation is achieved only by nonlinear optical processes known as two- or three-photon absorption. Such a process requires the nearly simultaneous participation of two or three photons, respectively, the rate of which, in turn, is proportional to the square or the cube of the beam’s intensity. As a result of this nonlinear dependency, the activation only occurs near the focal point and is avoided along the beam before or after the focus. This increased localisation compared to linear absorption is the key to the excellent resolutions obtained with this method and has the same physical basis as in multiphoton microscopy \cite{29,30,31}.

Furthermore, the lateral resolution, namely, the smallest lateral size of the material that can be polymerised, is effectively set by the diffraction limit of the multiphoton wavelength rather than that of the exciting laser. This corresponds to half or a third of the excitation wavelength for two- and three-photon processes, respectively. Since the nonlinear absorption process requires extremely high peak powers, ultrafast lasers producing femtosecond pulses are used. It is only this method where printing resolutions are indeed low enough to be limited by optical diffraction. The current state of the art corresponds to a repeatably achievable lateral resolution of $\sim \!\!150$ nm, although even 50 nm features have been reported \cite{8}.

\subsection{Subtractive Manufacturing}

Subtractive manufacturing, where the desired object is obtained by removing unneeded parts from a larger object, remains the dominant manufacturing method when the required number of units of the shape is too low for moulding, casting, and similar mechanisms to be impractical. In addition to various mechanical techniques, laser processing of materials is ubiquitous. An incredibly diverse set of industrial manufacturing processes, notably cutting, drilling, marking, metals, glass, plastics, ceramics, semiconductors, are switching from more mechanical techniques to lasers due to their greater precision, freedom from wear, and ease of automation. 

Lasers can be grouped according to their temporal characteristics, namely, continuous-wave (CW), nanosecond-pulsed, and ultrafast (pulses $<\!\!10$ ps). The laser-material interaction process is purely thermal for CW lasers. With nanosecond pulses, the laser power during a pulse is orders of magnitude higher than the average power. Material removal is through a form of vaporisation, known as ablation. Laser energy is better localised, but the heat-affected zone, where temperatures reach melting levels, is still 0.1-1 mm. The interaction of ultrashort pulses is markedly different. Energy delivery is so fast that the electrons, to which the laser energy is first coupled, cannot reach thermal equilibrium even with their atoms during the pulse. Often, there is no discernible heat zone; ultrafast ablation offers tangible benefits in quality and precision.

Due to their low costs, CW lasers are the most commonly used lasers in subtractive manufacturing. This approach’s typical precision and resolutions are in the 0.1-1 mm range. Nanosecond lasers offer better resolutions due to better containment of thermal effects but rarely get better than the 10-100 $\mu$mm range. Ultrafast lasers provide the greatest precision in micromachining, where structures are as fine as $\sim$1 $\mu$m can be controllably and repeatably removed, although 10 $\mu$m is more typical. The smallest structures obtained without using near-field effects are in the range of several 100 nm.

A significant advantage of subtractive techniques is that virtually any material can be processed with a matching technique. It is not that there are no limitations. For example, CW lasers are effective only on sufficiently absorbing materials at their wavelengths, and mechanical drilling of brittle materials is tricky, but there is a suitable technique available. In particular, ultrafast micromachining is nearly universal, from metals to semiconductors, glass to plastics and even biological tissue. This is so because the process of ultrafast ablation is mainly independent of the linear optical properties of the material, allowing even transparent materials to be removed. Also, the nearly complete avoidance of heating means that even thermally sensitive materials can be processed, albeit at low speeds. However, subtractive techniques lack the flexibility of additive manufacturing in creating arbitrarily shaped structures. A line of sight is needed for access, and complex shapes with voids may not be permissible. Furthermore, subtractive processing involves, by definition, the removal of material, which is often discarded. As such, it is highly wasteful.

\subsection{The  {\em Omni-modality} Approach: Merging of the Additive and Subtractive Manufacturing}

\begin{figure}[t]
\includegraphics[width=\textwidth]{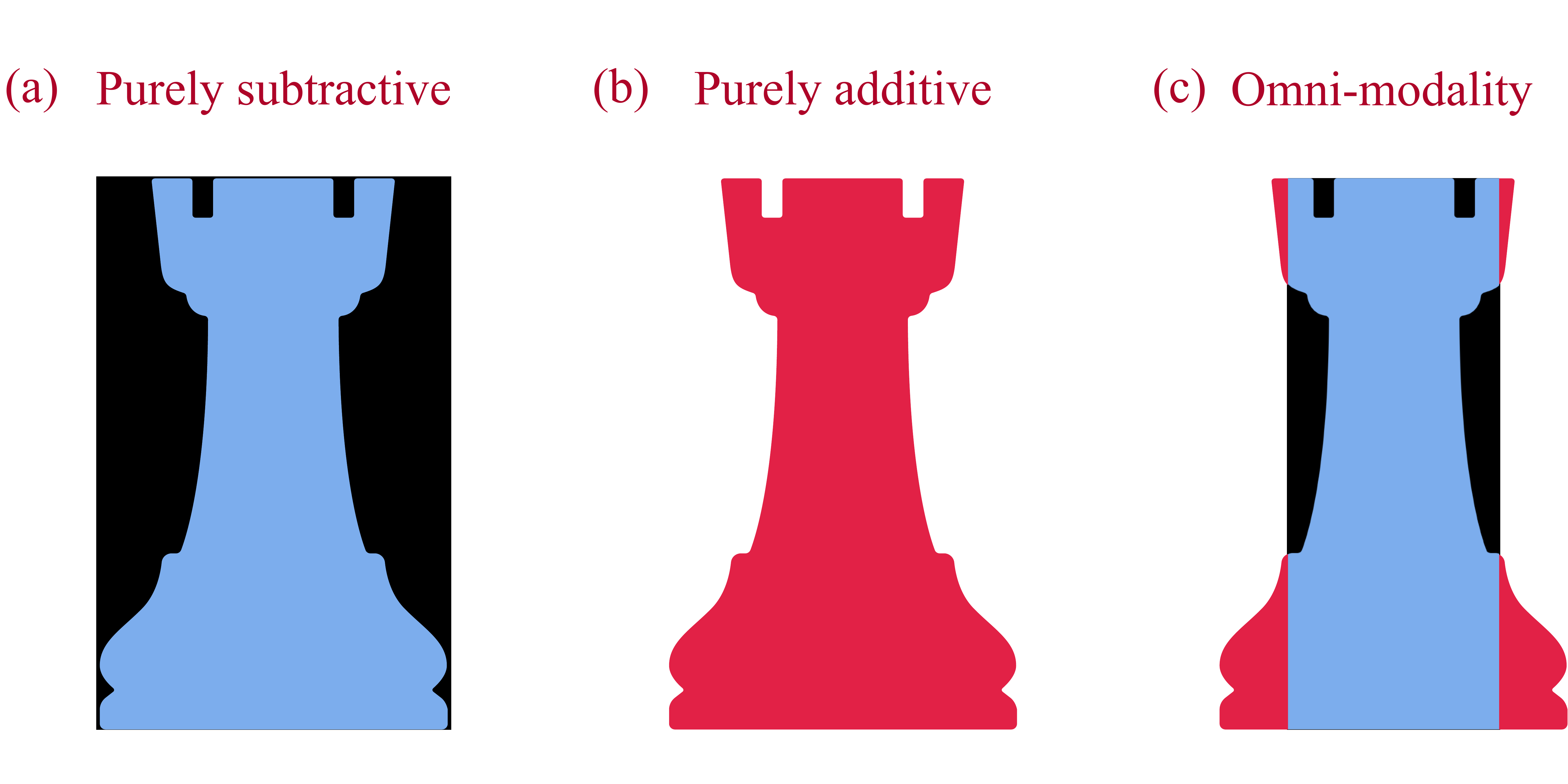}
\caption{(a) In subtractive manufacturing, a 3D object is created by removing it from an encompassing block. (b) Purely additive manufacturing starts from scratch, layer-by-layer printing the desired object by combining microscopic-sized parts. A depiction of the {\em Omni-modality}approach to laser-based manufacturing that seamlessly integrates additive and subtractive manufacturing around a single laser: (c) We start from an intermediate-sized block. Then, certain parts are removed, and certain features are added by 3D printing. The size of the original block is freely chosen to minimise the consumption of resources, including energy and time.}
\label{fig:omni}       
\end{figure}

Given the relative advantages and disadvantages of additive and subtractive techniques, we propose to combine them through the use of a single, versatile laser, which exploits the collective ultrafast pulse-matter interactions, and is capable of efficiently performing both techniques. We refer to this approach as the {\em Omni-modality} approach. One major reason is to reduce manufacturing time and material usage. The benefit of the combined use is a result of geometry in this case. As illustrated in fig. \ref{fig:omni}, a given object that needs to be created is often of irregular geometry, such that starting from a larger volume into which the intended shape fits requires the removal of substantial volumes (fig. \ref{fig:omni}a). In contrast, a purely additive approach requires printing the entire object, which is likely to be time-consuming (fig. \ref{fig:omni}b). Therefore, a logical alternative is to start from an optimally sized block, from which relatively little material has to be removed, and a similarly small amount of additional structure needs to be printed (fig. \ref{fig:omni}c). Moreover, the combined application of both techniques has various synergistic benefits that transcend increased speed and reduced waste. For instance, the final surface quality achieved with the additive techniques requires post-treatment, which could be improved \emph{in situ} using the flexibility to use ablation, melting and resolidification, or surface structuring, in any order, according to the need. The latter can include texturing for functionalisation. For example, self-organised patterns\cite{32} may be created on a surface to control its wettability, tribological \cite{33,34}, and biological \cite{35} properties.

\subsection{An {\em Omni-modality} Laser to Exploit the Collective Ultrafast Pulse-matter Interactions}

The {\em Omni-modality} approach is already achievable, but it would require the integration of multiple lasers, significantly increasing the complexity. This is so because lasers used for additive and subtractive processes, particularly ultrafast micromachining, are entirely different from those used for additive manufacturing. Based on our recent invention of the ablation-cooled laser-material removal technique \cite{27}, it is now possible to use the same laser for a wide variety of laser-matter interactions. Stated simply, the interactions for additive processes are based on thermal effects, except for the polymerisation method. At the same time, ultrafast micromachining works best in the so-called non-thermal mode, which means heating the nearby areas is avoided. From a physics point of view, the key feature of the ablation-cooled regime is that many, typically thousands of pulses, interact and ablate the material collectively. By judiciously exploiting the \emph{collective} nature of the process, it is possible to alter the interaction physics. Before clarifying this, we will briefly discuss ultrafast ablation and the ablation-cooled regime. 

Ultrafast ablation is the most precise mechanics of material removal using a laser beam. The electrons absorb a laser beam’s energy first, which transfers to their atoms within a few picoseconds. Femtosecond pulse-material interactions are known as non-thermal because the electrons and the atoms reach equilibrium after the pulse ends. In reality, the terminology of non-thermal is not accurate, as a thermal process does occur at the point of light absorption. A better expression would be non-thermal-equilibrium as the aspect that truly distinguishes ultrafast ablation from longer-pulsed or CW lasers is that the electron and atom temperatures are far from thermal equilibrium with each other during the pulse. Regardless of this sidenote on terminology, ultrafast ablation allows for the best localization of the laser’s effect. Surrounding regions of the material that are as close as a few micrometres can be prevented from heating substantially as the part of the material at focus is ablated. 

\begin{figure}[t]
\includegraphics[width=\textwidth]{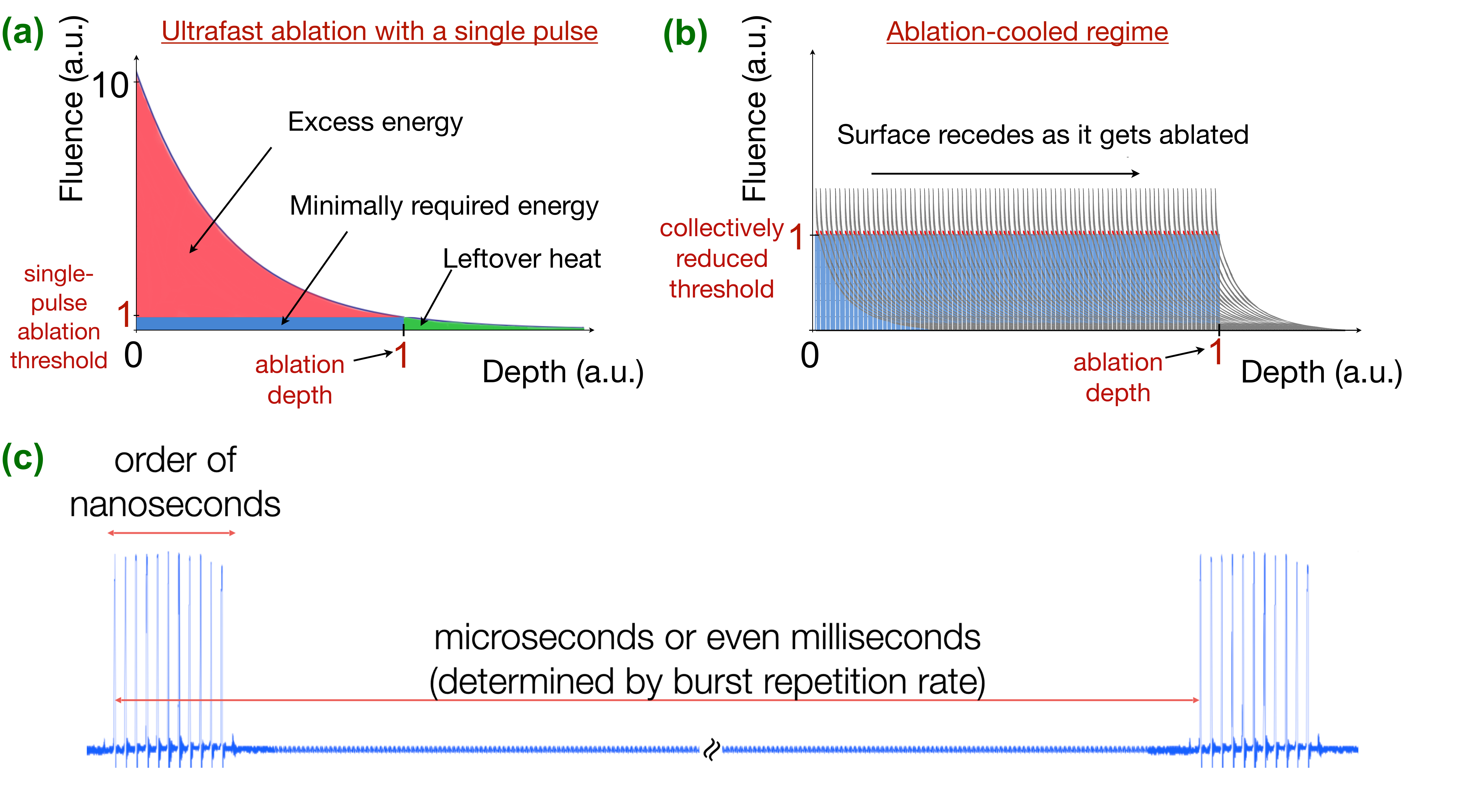}
\caption{(a) Depiction of energy absorption during ultrafast ablation with a single pulse, which avoids collateral heat damage but suffers from low efficiency. (b) In contrast, in the ablation-cooled regime, numerous pulses achieve higher efficiency through collective ablation, and the total energy under the many pulses is closer to the minimum energy within the blue-shaded area. (c) The temporal profile of a burst train is shown. The burst width is commonly 10-100 ns, and the time between successive bursts is given by the inverse of the burst repetition rate. This figure depicts a burst containing only ten pulses for clarity. Typically, we use bursts containing thousands of individual pulses and the repetition rate within each burst is up to 100 GHz..}
\label{fig:ultrafast}       
\end{figure}

Despite the advantages, ultrafast lasers remain a niche technology compared to the nanosecond and CW lasers despite their supreme precision because ultrafast micromachining is slow and inefficient. The fundamental reason for the low efficiency and speed is the Beer-Lambert law, which stipulates the exponential absorption of light (fig. \ref{fig:ultrafast}a). Ablation occurs only where the absorbed energy exceeds a threshold. As a result, ablation depth increases logarithmically with pulse energy \cite{36}, limiting efficiency.

The exponential tail below the threshold becomes excess heat (fig. \ref{fig:ultrafast}a, green-shaded region). In the commonly used method of ultrafast ablation, the material is allowed to cool by diffusion of this excess heat to surrounding regions before the next pulse arrives. Typically, the repetition rate is in the kHz range and rarely reaches 1 MHz. Therefore, the time between the pulses ranges from a microsecond up to a millisecond. If the repetition rate is too high, heat accumulation occurs, and even though femtosecond pulses are used, damage to surrounding regions is observed. In practice, the peak power of the pulses is chosen to be well above the ablation threshold to minimize the fraction of energy that is left behind in the form of excess heat. However, the fixed exponential form of the absorption necessarily means that much of the absorbed energy becomes extra energy for the material, which already has received enough energy to be ablated in the form of extra kinetic energy for the particles to be ejected (fig. \ref{fig:ultrafast}a, red-shaded region). The resulting logarithmic scaling of the ablated amount with respect to the pulse energy drastically reduces the efficiency of the ablation. Therefore, the exponential form of the Beer-Lambert law limits the repetition rate, and the rate at which material can be removed without causing cumulative heating. 

The ablation-cooled laser-material removal method \cite{27,37} is qualitatively different because thousands of ultrafast pulses, which have extremely short time gaps between them, {\em collectively} ablate the material (fig. \ref{fig:ultrafast}b). We use repetition rates up in the multi-GHz range, up to 50 GHz. The delay between pulses is so short that the heat left behind each pulse cannot diffuse before the next pulse arrives, and the temperature at beam focus becomes a {\em nonlinear iterative} function of all the previous pulses. Excess heat is extracted during ablation to replace diffusion as the dominant heat transfer mechanism, which is the origin of the term  {\em ablation cooling}. We have reported $>\!\!10$-fold higher efficiency with up to 1000-times lower pulse energies. Each pulse is far below the threshold for ablating the material independently in this regime. 

\begin{figure}[t]
\includegraphics[width=\textwidth]{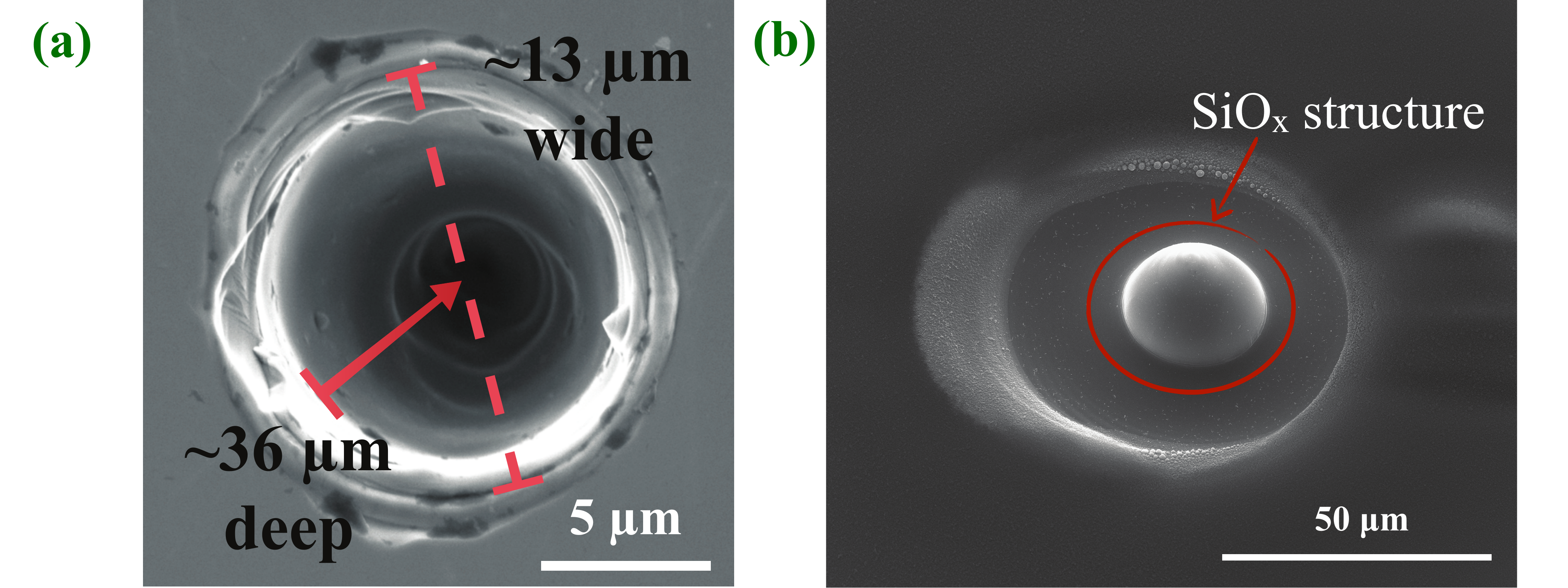}
\caption{Same sample, same laser, different parameters: (a) Top view of the crater created by a single burst as a record-efficient ultrafast ablation of crystalline Si. (b) The purely thermal activation of SiO$_{\rm x}$ growth is induced by keeping the GHz-train of pulses below the ablation threshold. The average power, spot size at focus, wavelength, and pulse duration were approximately the same in both cases. The main difference was the duty cycle of the bursts, which, in turn, changes the pulse energies. The laser is switched from one mode of operation to the other by purely electronic means.}
\label{fig:UniLase}     
\end{figure}

An important consequence of the collective ablation is that by changing certain pulse parameters, such as the repetition rate, the interaction can easily be taken below the ablation threshold without changing the average power applied. Once below the threshold, the collective interaction can be switched from a heat-accumulation-free ultrafast interaction to a purely heating mode, even though the pulses are still ultrafast. We can regard this mode as a quasi-CW mode. We have developed a versatile laser system that can electronically switch between different operational modes, namely, a purely thermal excitation mode and an ultrafast ablation mode are well suited to additive and subtractive manufacturing, respectively. The details of the laser setup are beyond the scope of this chapter and will be discussed elsewhere. This laser is based on a so-called burst-mode fibre laser, which was pioneered by one of us \cite{38,39}. Bursts refer to groups of high-repetition-rate pulses, followed by a long gap of no emission before another burst is produced (fig. \ref{fig:ultrafast}c). We have shown, experimentally, that the same laser can enact efficient ultrafast ablation and non-ablating thermal excitation by purely electronic reconfiguration of the temporal sequence of the pulses (fig. \ref{fig:UniLase}).

\begin{figure}[t]
\includegraphics[width=\textwidth]{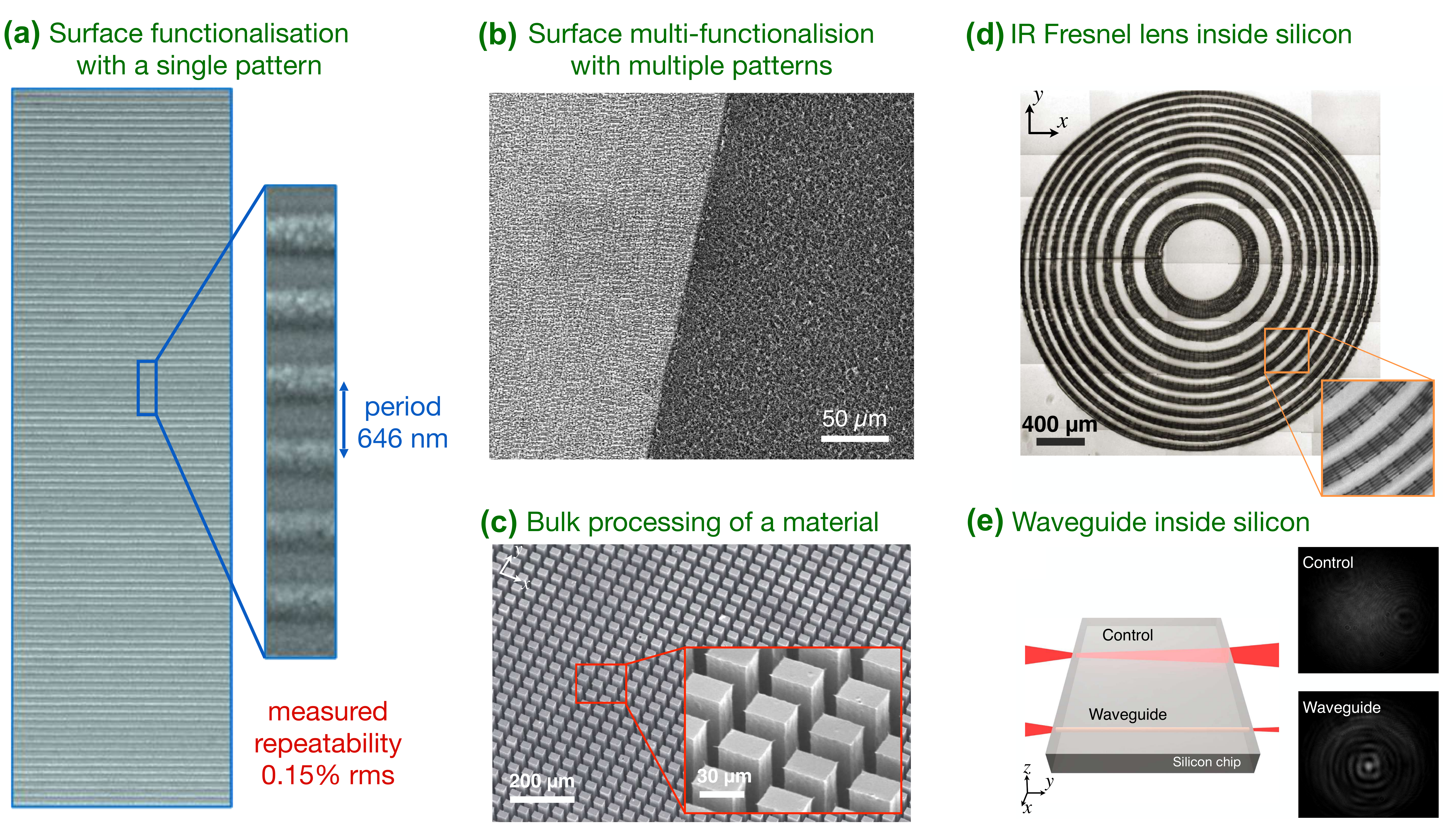}
\caption{Examples of surface and bulk functionalisation using 2D and 3D Nonlinear Laser Lithography technique. Scanning electron microscope images of highly uniform self-organised (a) TiO$_{\rm x}$ nanostructures on titanium \cite{32}, (b) SiO$_{\rm x}$ nanostructures on silicon, and (c) rectangular blocks carved inside silicon \cite{40}. Optical images of (d) an IR Fresnel lens and (e) a waveguide written deep inside silicon without processing its surfaces \cite{40}.}
\label{fig:NLL}     
\end{figure}

In addition to additive and subtractive manufacturing, a third type of laser processing is surface treatment or functionalisation. This can take multiple forms. The most straightforward applications include smoothening a rough surface either by selective melting and resolidification or ultrafast ablation. However, there are exciting additional possibilities that introduce functionalisation, namely, creating a surface texture or structure that achieves a predesigned function. Examples include self-organised patterns (fig. \ref{fig:NLL}a)32 that may be created on a surface to control its wettability (fig. \ref{fig:NLL}b), tribological \cite{33,34}, and biological \cite{35} properties. Furthermore, bulk materials can also be processed to functionalise materials \cite{40} for certain optical, electrical, and optoelectronic applications as can be seen from fig. \ref{fig:NLL}c-\ref{fig:NLL}e.

\section{Three Fundamental Challenges and Laser-driven Self-organization as a Remedy}
\label{sec:3}

Despite the rapid progress in 3D printing technologies, three fundamental challenges strongly limit the current paths towards approaching atomic resolution due to technological improvements. Instead, a radically different approach is necessary that circumvents these challenges. Before we propose the use of self-organisation as a possible path, we will discuss these three challenges.

\subsection{The \emph{Fat Fingers} Problem ($\lambda \gg a_0$)} 

As discussed in the preceding section, current approaches to creating arbitrarily shaped objects in a 3D printer are based on what is known as direct writing. Most of the laser-based techniques have micron-scale features sizes. The best resolution or the smallest voxel sizes are obtained using ultrafast lasers driving multi-photon-based polymerisation. In either case, the smallest feature sizes that can be written are ultimately limited by the wavelength of the light, and in the range of several microns for general-use techniques and can approach 100 nm in the case of two-photon polymerization or laser ablation with very tight focusing. Specific workarounds have been proposed, using plasmonics \cite{41}, metamaterials \cite{42}, or implementation of super-resolution microscopy techniques \cite{43,44}. Still, they each introduce new and stringent limitations \cite{45}, and the improvements are far from what is needed to approach the atomic scale. The wavelength of light imposes a fundamental barrier that is about several orders of magnitude distant from the inter-atomic distances. We succinctly summarise the fundamental origins of this limitation by the inequality, $\lambda \gg a_0$, where $\lambda$ is the wavelength of the laser and $a_0$ is the Bohr radius that sets the order of magnitude scale for the inter-atomic distances.

\subsection{The \emph{Explosion of Complexity} Problem} 

It is conceivable that the {\em Fat Fingers Problem} will be overcome in the future, for example, with the development of practical x-ray lasers or through the use of electron beams instead of lasers. That possibility prompts us to consider the formidable limitation of the {\em Explosion of Complexity}: As the feature size is scaled down, the number of points that must be processed increases cubically for 3D structures. Consider an object of 1 mm$^3$ in size, a rather modest size. At present, carving out or additively building such a structure at 1 $\mu$m resolution is possible but requires an order of 10$^9$ steps. Approaching the atomic scale, order of 1 $\mathrm{\mathring{A}}$, increases the number of steps to 10$^{21}$, which is in the realm of science fiction. At a rate of a million steps per second, the point-by-point processing of a 1-mm$^3$ structure at atomic resolution would take millions of years. Even if massive parallelisation and tremendous increases in the speed of processing each point can be anticipated, reaching the capability to process such astronomical numbers of points unlikely to become feasible within the foreseeable future. Equally importantly, the necessity of point-by-point processing of each atom is extremely questionable. One evident reason is that even the spatially most complex materials, such as biological tissue, harbour large amounts of repetition and redundancy. Moreover, the third challenge that leads us to question the utility of atom-by-atom processing, even if it becomes feasible and practical.

\subsection{The {\em Mischief of Fluctuations}}

The IBM demonstration was achieved at the cryogenic temperature of 4 K because as temperature increases, the random motion of atoms becomes increasingly strong as temperatures close to room temperature are approached. Therefore, a scaling down of the direct writing approach would have to confront increasingly strong fluctuations. This problem is exacerbated by the fact that the most straightforward solution to the limitations of diffraction is to use lasers with shorter wavelengths, even x-rays to reach the atomic scale. However, this means increasing photon energies, which would transfer even more energy to the atoms, increasing their kinetic motion, not to mention that the atoms would be ionised. Therefore, any method that aims to achieve a particular organisation at the atomic scale, has to be compatible with the inevitably thermal motion as the atoms are directed toward a particular atomic structure. 

Given these three strong limitations that arise fundamentally from the physics at play, it is difficult to believe that the currently dominant approach to 3D printing and micromachining, namely, direct writing, can reach the atomic scale, and remain practical, in the foreseeable future, if ever. These difficulties motivate the central thesis of this Chapter that self-organisation may be a viable route toward achieving control over the atomic structure.

\subsection{Laser-driven Self-organization as a Remedy}

\begin{figure}[t]
\includegraphics[width=\textwidth]{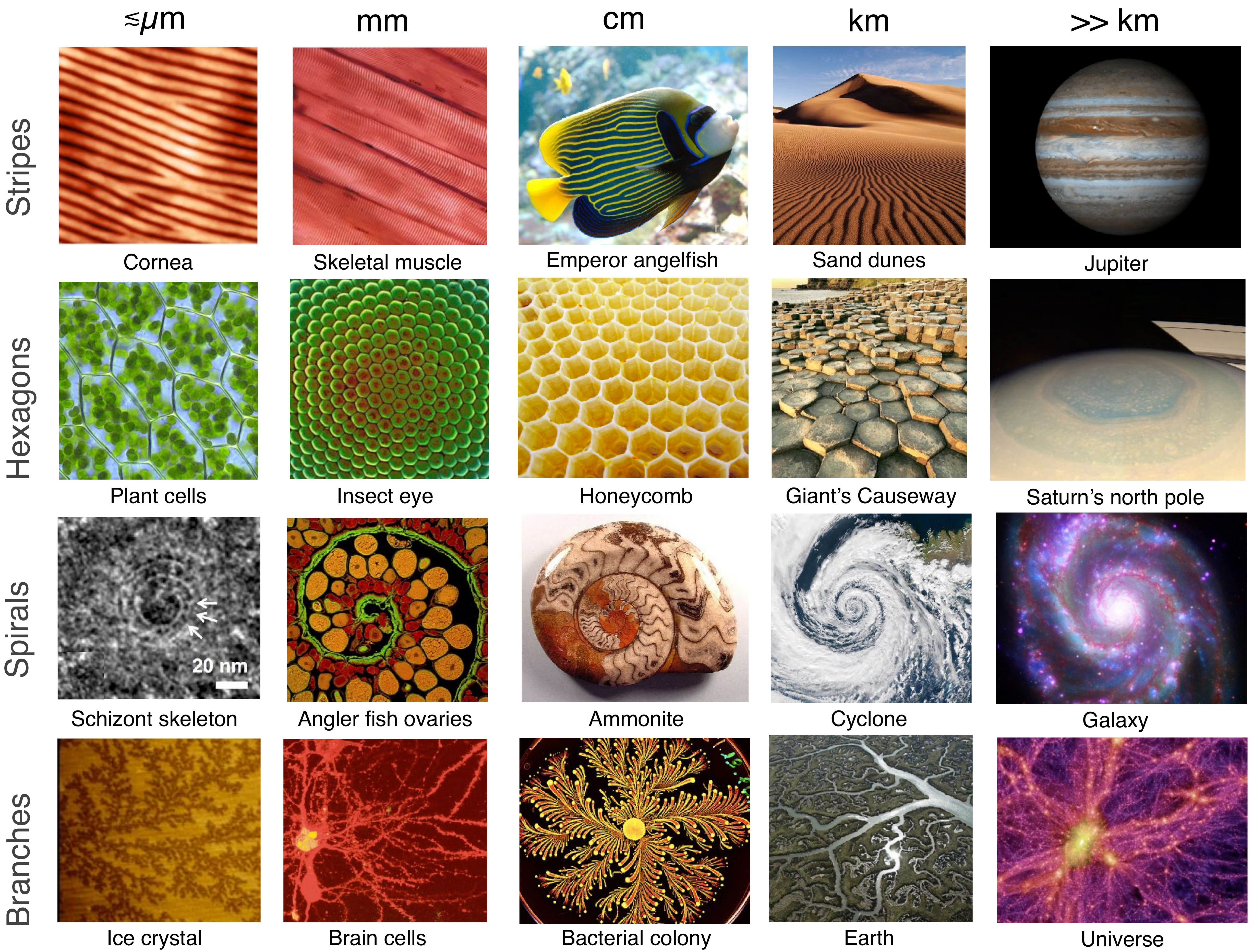}
\caption{Similar self-organised geometric patterns are observed across a wide range of spatial scales: Cornea \cite{50}, skeletal muscle (Image Credit: Michael Abbey), emperor angelfish (Image Credit: Kasper Hareskov Tygesen), sand dunes (Great Sand Dunes National Park; Image credit: Jack Brauer), Jupiter (Image Credit: NASA), plant cells (Image Credit: Kristian Peters), insect eye (Diamondback moth; Image credit: David Scharf), honeycomb (Image Credit: Cordelia Molloy), Giant’s Causeway (Image Credit: The Independent), Saturn’s north pole (Image credit: NASA), schizont skeleton \cite{51}, angler fish ovaries (Image Credit: Michael Gakuran), ammonite fossil (Image Credit: Unknown), cyclone (Image Credit: NASA), galaxy (The Whirlpool Galaxy; Image Credit: NASA), ice crystals \cite{52}, brain cells (Image credit: Mark Miller, Brandeis University), bacterial colony (Image Credit: Eshel Ben-Jacob), Earth (Salt marshes, Coto Do\~{n}ana National Park; Image Credit: Unknown), universe (Computer simulation by The Virgo Consortium; Image credit: The New York Times). This collage is inspired by Philip Ball’s book titled {\em Patterns in Nature}.
}
\label{fig:patterns}     
\end{figure}

Nature demonstrates self-organised pattern formation \cite{46,47,48} from the atomic to astronomic scales (fig. \ref{fig:patterns}), where the basic units of each physically different system assemble and form structure. Although they are often used interchangeably, and there are efforts to construct formal definitions \cite{49}, self-assembly and self-organisation fundamentally refer to the same phenomena of emergent complexity through spontaneous aggregation. Therefore, we will use self-organisation as an encompassing term that includes self-assembly.

The oldest publication documenting observation of self-organised pattern formation in a laboratory setting, as opposed to a natural occurrence, is that of convection cells in liquids, to our knowledge, and dates to 1901 \cite{31}. The first report of specifically laser-induced self-organised pattern formation is from 1965 \cite{53}. Today, self-organisation has been documented and studied in numerous laboratory systems. Since the earliest studies, self-organization has been primarily investigated for chemical systems \cite{54,55,56}, which has been later extended to other physical systems \cite{57,58,59,60,61,62}, and fields of engineering \cite{63,64} and social sciences  \cite{65}. There is vast literature on self-organization with beautiful examples. Unfortunately, it remains mainly a phenomenon observed when encountered, often with fascination. Its deliberate use to achieve technological functions is relatively rare, which we attribute to the lack of a general theoretical framework that allows the prediction of the conditions, such as a set of externally controllable parameters, that will result in a self-organised system to reach a pre-selected pattern and often one is required to laboriously search for the right condition of settings in a large parameter and phase space. Despite the enormous progress in self-organisation, the most severe unresolved difficulty from the perspective of a 3D atom printer is the specificity of self-organisation processes. For such a printer to be practical, it should offer near-universality or at least an extensive library of material compositions and atomic structures that can be assembled. Unfortunately, self-organisation literature is by and large material-, size-, shape-, and interaction-specific.

Virtually each demonstration of self-organisation results from carefully engineered inter-particle interactions, interactions between the particle system and its environment, and the external energy source, which diminishes hopes for constructing a general theoretical framework for self-organized systems. For instance, it is pretty common to engineer the interactions between colloidal particles by decorating them with (bio)chemicals (\emph{e.g.}, Janus particles \cite{66}, DNA origami \cite{67} techniques, and alike) or by promoting specific chemical bondings between certain molecules \cite{56}. It is also popular to engineer the chemical composition of the liquid solutions to direct the movements of such functionalised particles68. Other engineered interactions involve particles that can only respond to a specific external energy source, such as a nanoparticle swarm that moves only in a magnetic field69, or micron-sized robots that could only be controlled by the electrical signals emitted by an electronic device70. All these engineered interactions render the principles leading to self-organization specific to the materials or chemical used, external energy source, or experimental conditions.

Each demonstration of self-organisation resulted from carefully engineered inter-particle interactions, interactions between the particle system and its environment, and the external energy source, which prevents constructing a general theoretical framework for self-organized systems. For instance, it is pretty common to engineer the interactions between colloidal particles by decorating them with (bio)chemicals (\emph{e.g.}, Janus particles \cite{66}, DNA origami \cite{67} techniques, and alike) or by promoting specific chemical bondings between certain molecules \cite{56}. It is also popular to engineer the chemical composition of the liquid solutions to direct the movements of such functionalised particles \cite{68}. Other engineered interactions involve particles that can only respond to a specific external energy source, such as a nanoparticle swarm that moves only in the presence of a magnetic field \cite{69}, or micron-sized robots that could only be controlled by the electrical signals emitted by an electronic device \cite{70}. All these engineered interactions render the principles leading to self-organization specific to the materials or chemical used, kind of the external energy source, or to experimental conditions. 

As illustrated through the geometric patterns of fig. \ref{fig:patterns}, self-organization occurs over vastly different physical scales but yields similar features in these unrelated systems. A most pertinent question is whether we can harness this capacity to achieve a high degree of independence of directed self-organised systems from their specifics. This is a matter of great practical importance as it would address one of the most limiting aspects of the self-organised approach, namely that virtually every experiment has to be redesigned when the materials or the interactions involved are changed.

\begin{figure}[t]
\includegraphics[width=\textwidth]{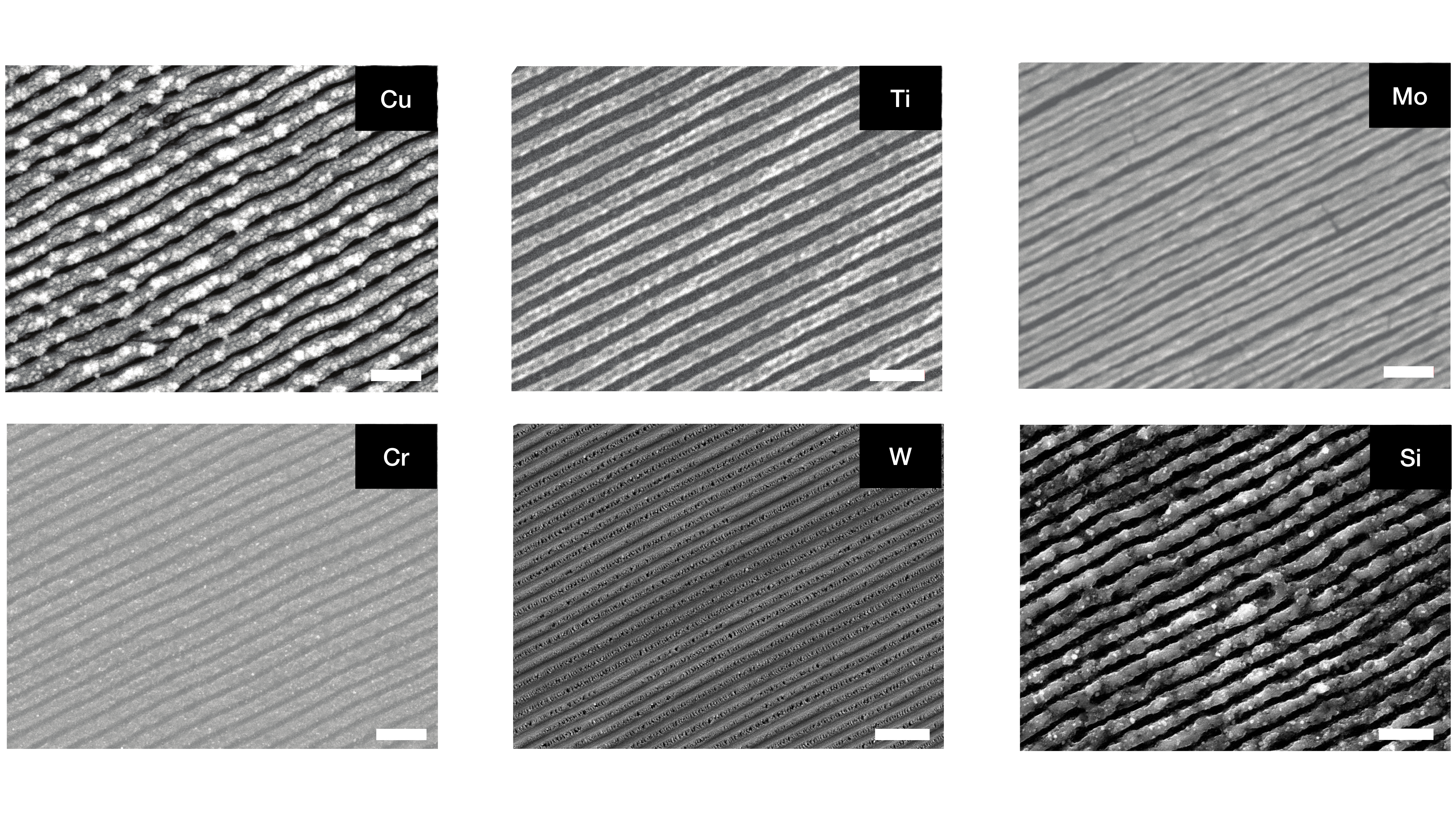}
\caption{The universality of nonlinear laser lithograph (NLL) \cite{32} is demonstrated on various materials.}
\label{fig:NLL_universality}     
\end{figure}

There is accumulating evidence that the answer may be affirmative, as demonstrated through both NLL  \cite{32} and universal dissipative self-assembly \cite{61} methods we introduced to the scientific literature. We have already confirmed that NLL is able to produce the same structures using Ti, Mo, W, Cr, Cu, Al, Si, GaAs, and glass, some of which can be found in fig. \ref{fig:NLL_universality}.

\begin{figure}[t]
\includegraphics[width=\textwidth]{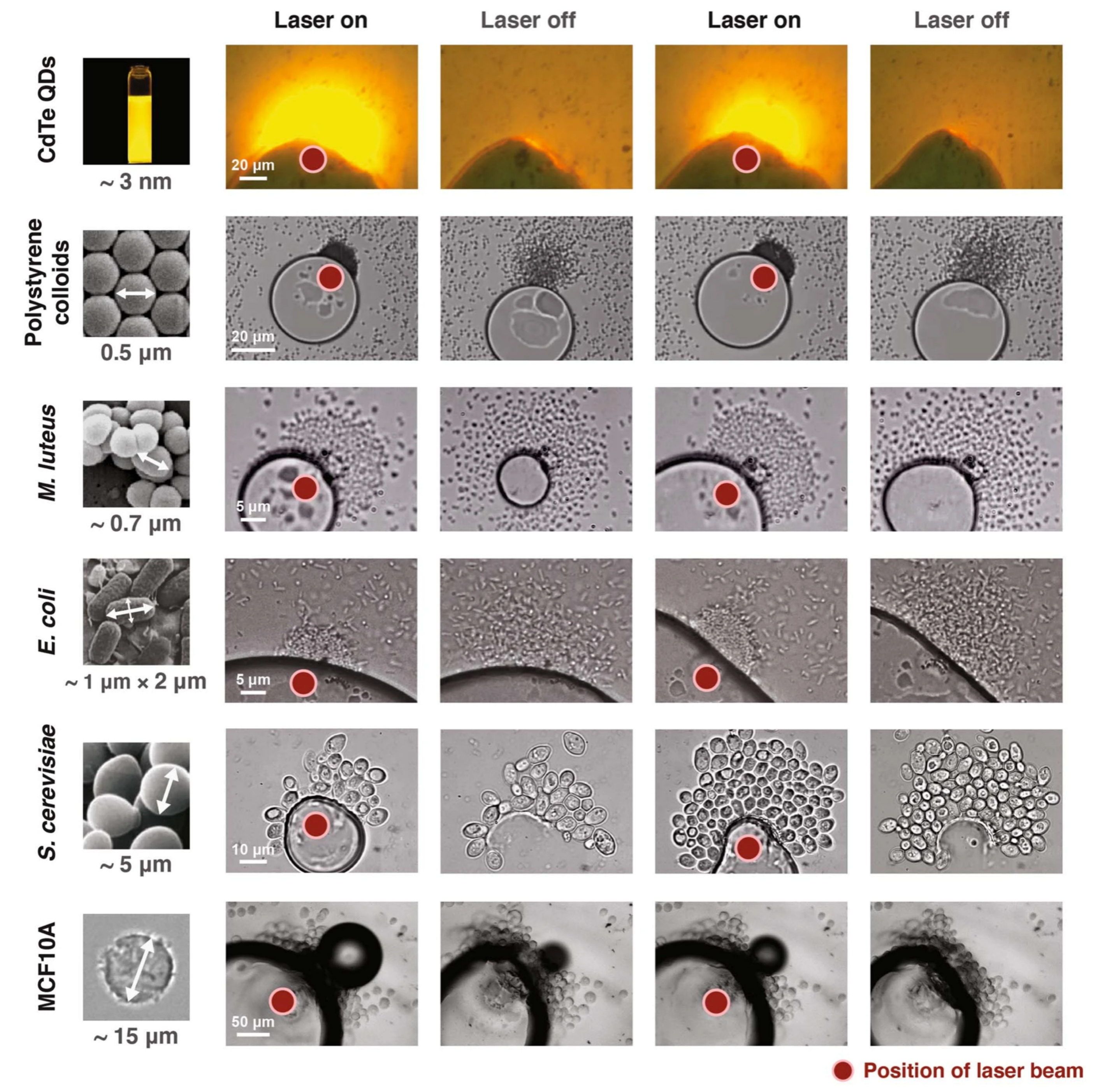}
\caption{Dissipative self-assembly universality demonstrated with various particles and organisms \cite{61}.}
\label{fig:DSA_universality}     
\end{figure}

Even more dramatically, we have achieved self-assembly of $\sim$3-nm-large CdTe quantum dots suspended in water (or various chemicals), 500-nm-large polystyrene spheres in water, ~0.7-µm-large soft spheres of {\em Micrococcus luteus} and $\sim$1-$\mu$m by $\sim$2-$\mu$m in size, rod-like {\em Escherichia coli} bacterial cells, $\sim$5-$\mu$m elliptical {\em Saccharomyces cerevisiae} yeast cells, and $\sim$15-$\mu$m-large MCF10A normal human breast cells, which obey the same physical rules (fig. \ref{fig:DSA_universality}) \cite{61}. In all of these examples, key experimental parameters, such as laser power, naturally need to be adapted to achieve optimal results in each case. Still, the same qualitative self-organised behaviour is observed, which suggests a high degree of universality and material independence. Setting aside the promising experimental demonstrations, these two methods and other model systems with dissimilar dynamics have to be further studied to formalise a general theoretical framework to create design recipes for an atom printer.

The following subsection provides arguments on how ultrafast lasers are outstanding candidates as model systems to study self-organization and as tools to drive it. We will also discuss how they help address the three fundamental challenges we identified at the beginning of this section.    

\subsection{The Promise of Ultrafast Lasers}

The (passive) modelocking of lasers is arguably the most successful application of human-made self-organisation. First demonstrated in 1964 \cite{71}, passive modelocking developed to underlie the science and technology of ultrafast optics with diverse applications, including precision material processing, optical imaging, frequency combs, and high-field physics, among many others. At its core, modelocking is the self-organised formation of ultrafast pulses in a laser cavity, which is achieved because the frequency modes of the laser cavity experience higher net laser gain when they mutually lock up in phase \cite{72}. Perhaps because of its success, self-organisation underlies the generation of ultrafast pulses through modelocking may be overlooked \cite{73}. This is a pity since the modelocking of lasers is an excellent model system for 1D self-organisation that can be described with a master equation introduced by Hermann A. Haus in 1991 \cite{74}. There is no fundamental limitation to studying the dynamics leading to the self-organisation of thousands of frequency modes in much more complex self-organising 2D and 3D systems, following the advent of spatiotemporal modelocking. We envision that such efforts may even build upon Haus’ master equation and arrive at a general theory of self-organisation.

Self-organisation is dissipative. From the perspective of nonlinear systems, Liouville’s theorem shows that conservative systems cannot support attractors to which an extended portion of the phase space decisively converges. Similarly, the sustained temporal evolution of a statistical system requires detailed balance to be broken; a self-organised process needs to be driven away from thermal equilibrium by an external energy flux accompanied by dissipation. Numerous mechanisms can introduce gradients in temperature, pressure, concentration, probabilities or any other parameter that is relevant to that particular system. In Nature, this source can be, for instance, the Sun, or tidal and geological heating, or chemical reactions, as in ATP synthesis by electrochemical gradients \cite{75}. In the laboratory, it can be electrical, optical, a plasma source (\emph{e.g.}, magnetrons in a vacuum system), chemical reactions, or a laser. Fundamentally, which energy source is used does not matter; the important thing is to create gradients in transition/interaction probabilities. Nevertheless, among these, lasers, particularly ultrafast lasers, stand out with a unique combination of advantages. 

\begin{figure}[t]
\includegraphics[width=\textwidth]{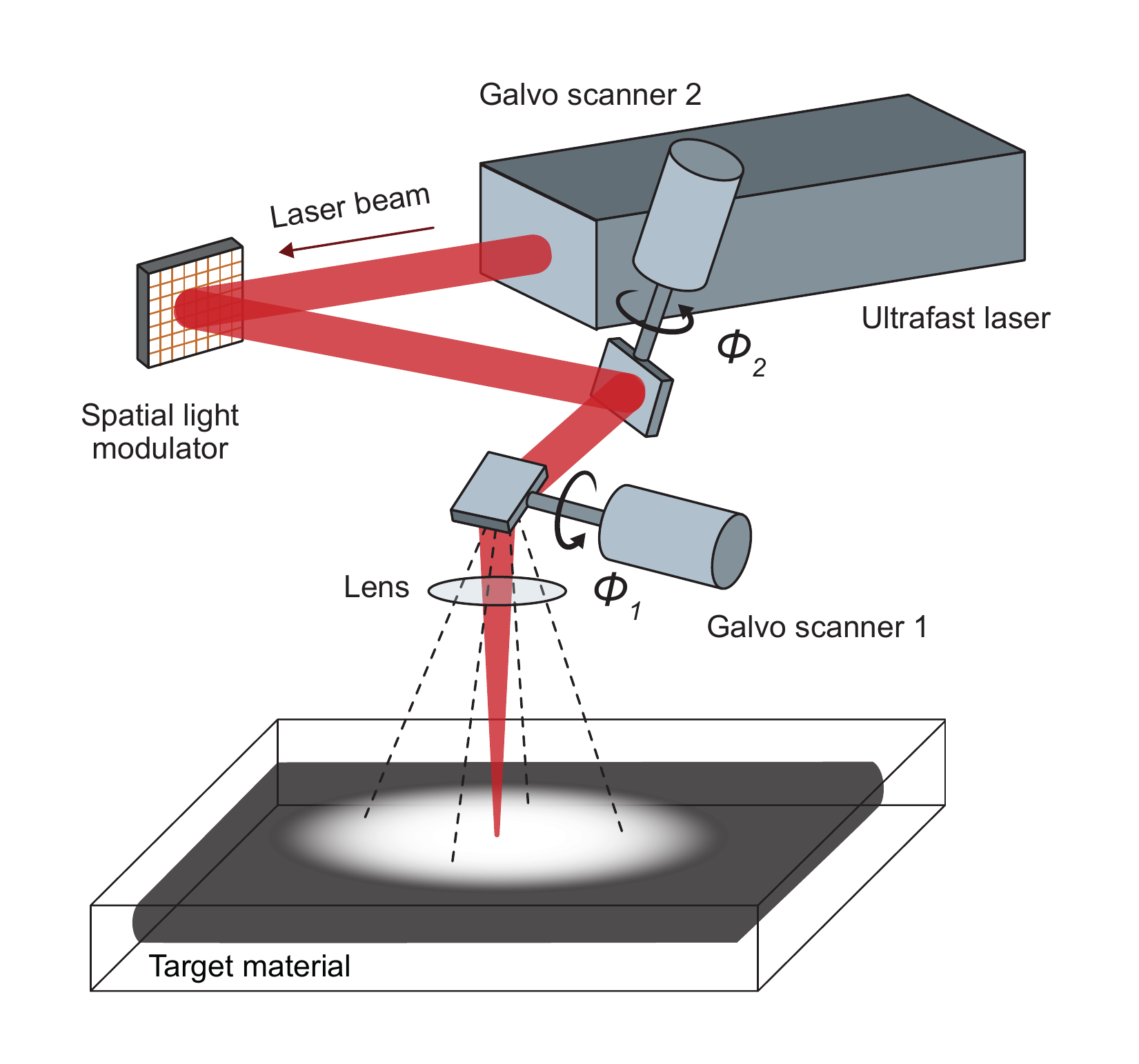}
\caption{The most common method of using an ultrafast laser coupled to a rapid beam scanner followed by a lens or objective that focuses the beam on the target material. Shown here are a two-axis galvanometric scanner that allows beam displacement within $\sim \!\!1$ms and a spatial light modulator that allows wavefront and beam shaping, as well as displacement.}
\label{fig:beam}     
\end{figure}

Ultrafast lasers can induce spatiotemporal thermal gradients on the materials they interact with. Single or multi-photon absorption at a material converts the energy of the pulses into nearly instantaneous local heating, which can be used to create highly configurable adjustable spatiotemporal thermal gradients. The material could be a metal, semiconductor, glass, ceramic, a polymer \cite{32}, colloidal solution of nanoparticles \cite{57,61} or living organisms \cite{61}, a transparent material \cite{40}, a thin film grown under vacuum \cite{76}, among other possibilities. Peak thermal gradients of $>\!\!10^7$ K/mm can momentarily be achieved using ultrashort pulses despite delivering only minuscule amounts of energy (nanojoules per pulse). Laser power, thus, the rate of energy delivery can be electronically adjusted in nanoseconds. The beam from a modelocked ultrafast laser can be shaped to produce any desired shape at focus, limited only by diffraction, using the millions of pixels of a spatial light modulator as actuators \cite{61}. The focussed beam can then be moved to any point on the surface or at any depth in the case of a transparent target within microseconds to milliseconds \cite{40}. Using these capabilities, one can create nearly arbitrarily sculptured, highly precise, and extraordinarily steep thermal gradients (fig. \ref{fig:beam}). 

Nevertheless, these advantages appear to have been underappreciated, and the overwhelming majority of their usage focused on ultrafast measurements of transient phenomena, excitation of multiphoton processes, \emph{e.g.}, for microscopy, or, even more commonly, to ablate materials \cite{77,78}. The last one is merely an extreme consequence of creating a large thermal gradient.

\subsection{Overcoming the \emph{Fat Fingers} Problem}

In self-organisation, the entire structure, from its finest features to the overall pattern, is determined by the nonlinear interactions. As such, there is no need for the energy source, say, the laser beam, to be localised at the same scale as the finest features to be created. Often, the pattern formation is driven by a thermal, chemical, another gradient, which is not even in the plane of the patterns. Consider the Rayleigh-B\`{e}nard convection cells, which are fluid patterns that form due to a uniform temperature difference along the direction perpendicular to the plane of the patterns \cite{46}.

Using well-controlled spatiotemporal thermal gradients to direct self-organisation provides a most interesting but hitherto unrecognised route towards beating the diffraction limit. The beam size is not relevant to the smallest features of the pattern. Still, it does limit the size of the pattern or the smallest distance over which one self-organised pattern can turn into a different one. In simpler terms, the diffraction limits the smallest spatial region within which the structure can be controlled independently, which limits the smallest voxel size. Even within the confines of this limitation, the smallest region, where a given self-organised structure is controllably created, can be substantially smaller. 

Is the diffraction limit entirely irrelevant when driving self-organisation with a laser beam? In self-organisation, the entire structure, from its finest features to the overall pattern, is determined by the nonlinear interactions. As such, there is no need for the energy source, say, the laser beam, to be localised at the same scale as the finest features. Often, the pattern formation is driven by a thermal, chemical, another gradient, which is not even in the plane of the patterns. Consider the Rayleigh-Bénard convection cells, which are fluid patterns that form as a result of a temperature difference along the direction perpendicular to the plane of the patterns \cite{28}.

The use of well-controlled spatiotemporal thermal gradients to direct self-organisation provides a most interesting, but hitherto unrecognised route towards beating the diffraction limit. As exemplified below, the beam size is not relevant to the smallest features of the pattern, but it does limit the size of the pattern or the smallest distance over which one self-organised pattern can turn into a different one. In simpler terms, the diffraction limits the smallest spatial region within which the structure can be controlled independently, which limits the smallest voxel size. Even within the confines of this limitation, the smallest region, where a given self-organised structure is controllably created, can be substantially smaller. 

\begin{figure}[t]
\includegraphics[width=\textwidth]{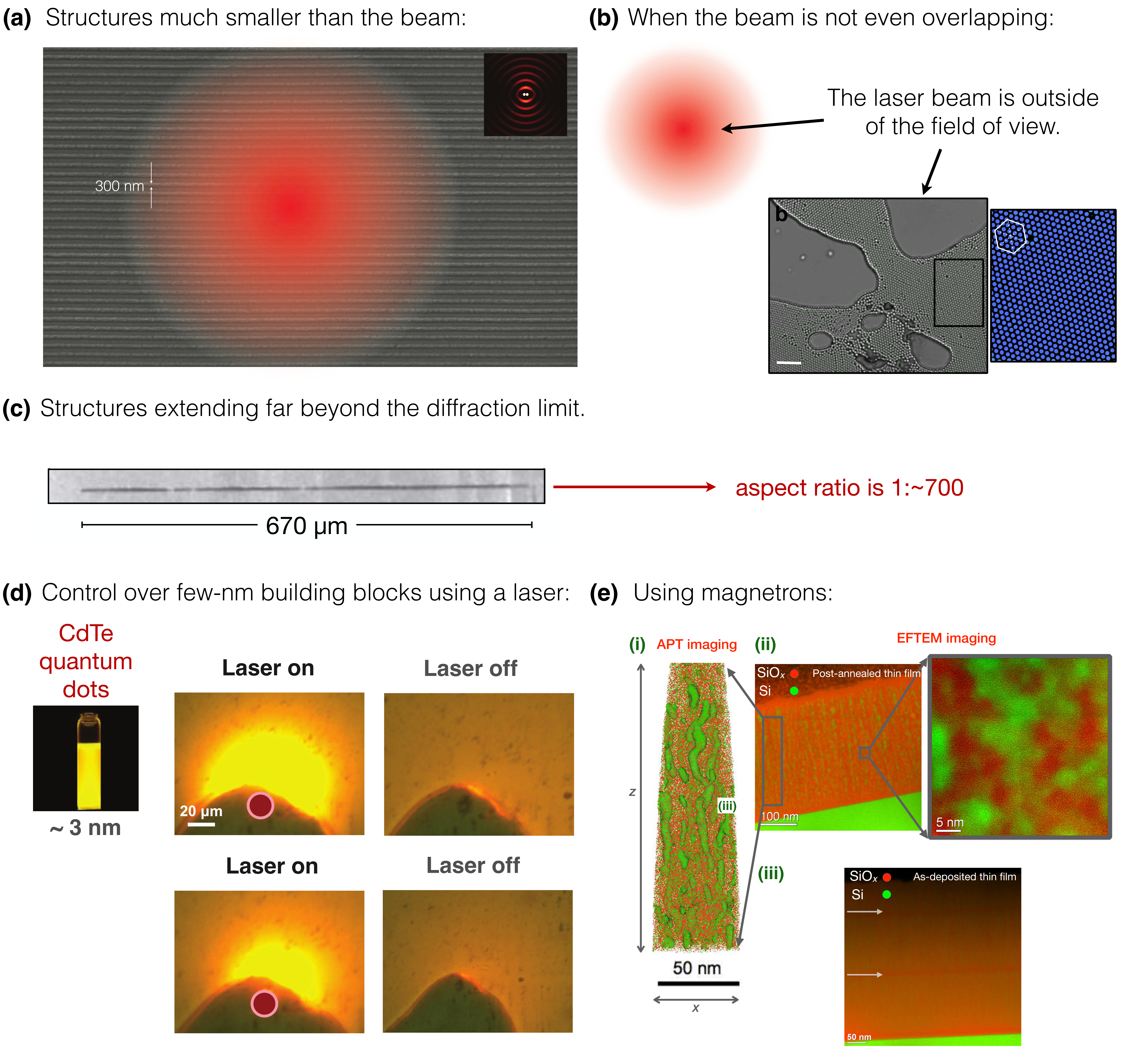}
\caption{A representative selection of recent experimental demonstrations showing fine control in the microscopic even few-nanometre range can be achieved without requiring energy flux with comparable foci. (a) Laser-induced periodic surface structures (LIPSS) with unprecedented uniformity and repeatability formed by scanning a Gaussian beam over the surface, known as nonlinear laser lithography (NLL) \cite{32}, is shown in comparison to the much larger beam used to create them. (b) A colloidal aggregate is formed by the dissipative self-assembly method shown \cite{57}, where the laser beam driving the process does not even necessarily overlap with the aggregate. (c) In-chip self-organized line formation \emph{via} NLL forming structures far beyond the diffraction limit \cite{40}. (d) Even $\sim\!\!3$ nm-large quantum dots can be collected with the dissipative self-assembly method61 shown in (b), where in both cases, the features of the self-organised pattern are determined by the interaction dynamics and not the size of the laser beam. (e) Even crude spatiotemporal thermal gradients can lead to controlled atom-by-atom formation of a complex structure, an anisotropic random network of Si quantum dots. The network is anisotropic at the mesoscopic scale but isotropic in the nanoscale of $\sim\!\!10$ nm, confirmed by atom probe tomography (APT) of structure and independently by energy-filtered transmission electron microscopy (EFTEM). A control experiment -c- proves that the origin of the multiscale structure is the spatial thermal gradient \cite{80}.}
\label{fig:size_no_matter}     
\end{figure}

Consider a dynamic process that leads to a particular self-organised material form only over a finite temperature range, $\Delta T = T_{\rm max} - T_{\rm min}$. Diffraction limits the smallest spot that a laser beam can be focussed to, known as the Abbe limit \cite{79}. The thermal gradient generated by the absorption of a short laser pulse is given by the ratio of the temperature difference created by the absorption of a laser pulse to the beam spot size. Diffusion reduces this gradient over time, but in the time scales relevant to ultrashort pulses and processes, the thermal gradients that can be reached with a diffraction limited beam can be nearly as much as $(T_{\rm c} - T_{\rm o})/d \approx T_{\rm c}/d$, where $T_{\rm c}$ is the critical temperature beyond which ablation occurs (in the range of $10^4$ K), To is the ambient temperature (say, $\sim\!\!300$ K) and $d = \lambda/(2NA)$ is the Abbe limit (for green light, $\sim\!\!250$ nm), where NA is the numerical aperture. For a sufficiently fast process for which we can ignore the effects of diffusion, our hypothetical self-organised structure can be confined within a spatial region of size $d \Delta T/T_{\rm c}$. Since it is possible that $\Delta T/T_{\rm c}< 1$, the fundamental limits to spatial localisation of self-organised material forms are not strongly limited by diffraction, and feature sizes considerably smaller than the diffraction limit are, in theory, achievable. It should be emphasised that creating a structure confined to a slice along the slope of a diffraction limited beam is not independently controllable and will apply to specialised usage scenarios. It is safe to consider the diffraction limit as the smallest voxel size for all intents and purposes. However, as discussed below, the self-organised structure within each voxel is not limited by diffraction. Substantial control over the atomic structure appears to be entirely possible. In fig. \ref{fig:size_no_matter} we present examples from our recent work, conducted in this direction.

In Nonlinear Laser Lithography (NLL) \cite{32,40} and numerous related works, the laser beam is homogenous and much larger (by 10-100 times) than the periodic structures that are self-organised, which are, in fact, smaller than even the laser’s wavelength (fig. \ref{fig:size_no_matter}a and \ref{fig:size_no_matter}c). This decoupling of the smallest features and the spatial localisation of the laser beam is even more dramatic in the case of dissipative colloidal self-assembly \cite{57,61} (fig. \ref{fig:size_no_matter}b and \ref{fig:size_no_matter}d). In this case, the self-organised aggregate is not even directly underneath the laser beam but typically displaced considerably, confined within a cavity bubble. The laser-induced Marangoni flows give rise to the aggregation process through drag forces \cite{57,61}. Furthermore, since the drag force is equally present for small and large particles, we have shown that the same self-assembly process works with CdTe quantum dots, as small as 3 nm, which are nearly four orders of magnitude smaller than the size of the laser beam. The laser-based control in the examples given above involves thermal gradients with a relatively high spatiotemporal resolution, whereby it was possible to achieve control over the self-organised patterns. However, it is possible to influence the atomic scale arrangement even by using thermal gradients with low spatial resolution along one direction (fig. \ref{fig:size_no_matter}e) \cite{80}.

\begin{figure}[t]
\includegraphics[width=\textwidth]{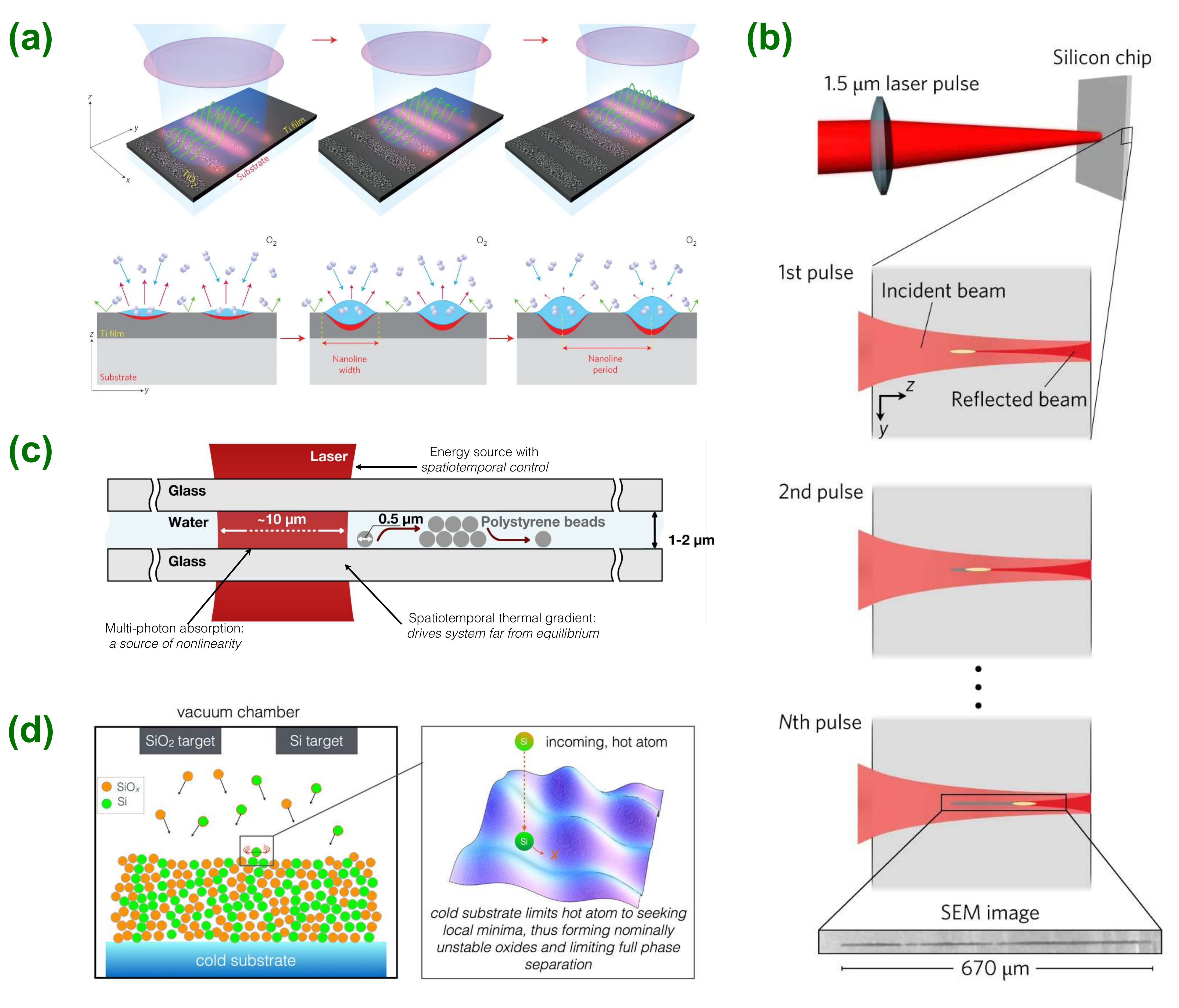}
\caption{Schematic representations of methods that lead to self-organised structures shown in fig. 10. (a) 2D-NLL \cite{32}, (b) 3D-NLL \cite{40}, (c) dissipative self-assembly of colloids \cite{57,61}, (d) self-assembly of Si atoms in a vacuum chamber \cite{80}.}
\label{fig:NLL_DSA_nano}     
\end{figure}

It is often the case that a steady state of a self-organised system is born out of fluctuations that grow through a positive feedback mechanism (also known as amplified fluctuations \cite{81}), which is counter-balanced by a negative feedback mechanism that is initially non-existent or negligibly weak, but becomes important once the amplitude characterizing this state grows beyond a certain size. Thus, a steady state is a dynamic balance that continues to exist in the presence of fluctuations. Luck or serendipity has nothing to do with achieving these results as the self-organisation in these different systems is guided by controlling the intrinsic feedback mechanisms \emph{via} a few external parameters. The key is to identify the positive and negative feedback loops and to which control parameter(s) and in what way they are responding. Here, a positive feedback loop refers to the situation where a certain change in the initial status of a system induces further changes, often leading to an exponential growth of the tiniest variation that typically arises from fluctuations, noise. An especially well-known example is the onset of coherent laser oscillation starting from spontaneous emission in an optical cavity that provides feedback. Eventually, a saturation mechanism (a negative feedback loop) is induced, slowing down the growth rate. A negative feedback loop acts to oppose a change, and it is known as an inhibitory process in chemistry and biology. As such, positive feedback is regarded as a source of instability and is desperately avoided. Negative feedback is the predominant form of external feedback applied in engineering applications, often to reduce noise, but our interest here is when both positive and negative feedback is internal and intrinsic to the system’s dynamics. Figure \ref{fig:NLL_DSA_nano} shows the depictions of the processes that resulted in the self-organisation of structures shown in fig. \ref{fig:NLL_DSA_nano}. Below, we will describe the intrinsic feedback mechanisms guiding self-organisation. 

The main reason for picking these model systems as case studies is that they are just about as different as they can be from each other. At the same time, self-organisation is born out of amplified fluctuations and guided by their intrinsic feedback mechanisms. Another reason for choosing them is that each is more conveniently described by one of the two major classes of theoretical models commonly used to describe pattern formation. These are, namely, models based on an interaction kernel (involving integro–differential or –difference equations) or a generalized reaction-diffusion model (a Turing-like instability, involving partial differential equations) \cite{82}. We should stress that these formalisms are interchangeable and do not represent any fundamental differences. 

Nonlinear laser lithography (NLL) judiciously exploited intrinsic feedback mechanisms to achieve self-organised structures of extremely periodic structures with uniformity (as characterized by Allan deviation) of 0.1 nm for structures with a period of 700 nm measured over long distances (extending over 100’s of $\mu$m) \cite{32}. The top row of fig. \ref{fig:NLL_DSA_nano}a shows the interaction of femtosecond pulses with the material surface. Each pulse scatters from the pre-existing roughness on the surface or by the nanostructures if already formed. The scattered light, propagating as surface waves, interferes with each other. At the points of constructive interference, the intensity is higher, which modifies the surface. The modified surface scatters the lights of the next pulse differently, which reinforces further modification, constituting a positive feedback mechanism. The bottom row depicts the negative feedback that regulates the growth of the structures. Atmospheric oxygen cannot penetrate through the growing oxide structure, limiting the reaction and preventing further growth of the structure. In the 3D-NLL case (fig. \ref{fig:NLL_DSA_nano}b), a self-focusing laser beam modifies the interior of the silicon chip without leaving any damage to its surfaces (positive feedback loop). The refractive index of the modified regions changes, which shifts the focal point of the iterative pulses and limits further growth of the self-organised structures (negative feedback). 

Similar to the NLL case, dissipative self-assembly of colloids (fig. \ref{fig:NLL_DSA_nano}c) benefits from the feedback mechanisms, where ultrafast pulses create nonlinear thermal gradients, which, together with the surface effects, give rise to Marangoni flows \cite{57,61}. A positive feedback mechanism occurs between the fluid flows that drag the colloidal particles and promote aggregation of the particles. The growth of the aggregates is counterbalanced by strong Brownian motion, which functions as a limiting force to the growing aggregates. In the vacuum deposition case shown in fig. \ref{fig:NLL_DSA_nano}d, deposition of Si and SiO$_{\rm 2}$ from two targets that were activated by magnetrons, which “rained down” on a substrate at the bottom of a vacuum chamber. A spatial gradient formed in the vertical (growth) direction of the substrate because of the steep temperature difference between the “hot” atoms and molecules that were energised by the magnetrons, and the cold substrate, which was kept at room temperature. By controlling the power of the magnetrons for each of the targets, the energies of the Si atoms and the SiO$_{\rm 2}$ molecules were controlled. Here, we promoted nanocrystal growth in the vertical direction to decrease the onset of percolation for achieving a network structure where individual quantum dots are connected by Si bridges (positive feedback). We also limited full phase separation of metastable oxides into Si and SiO$_{\rm 2}$ to avoid further growth of QD diameters (negative feedback).

In all these examples, the fundamental reason why self-organisation works so perfectly is that the feedback mechanisms driving the processes are nonlocal. Nonlocality refers to the situation where the feedback mechanisms acting at a particular point in space depend not only on the characteristics of that point but also over an extended region surrounding that point. This dependence can be described most conveniently in an integral form using a kernel that characterizes the nonlocal nature of the interaction, not only for the examples we have shown in fig. \ref{fig:size_no_matter} but also for diverse systems, ranging from self-organized vegetation patterns to spin glass systems.

\subsection{Overcoming the Explosion of Complexity}

Unlike the case of direct writing methods discussed in Section \ref{sec:2}, when thousands of particles \cite{57,61}, or even an astronomical number of atoms \cite{32,40,80} self-organise into a predetermined pattern, there is no need to control individual particles. The concept of self-organisation relies on the possibility that the system of interest, which typically has numerous degrees of freedom, can somehow be directed using just a handful of control parameters. This becomes possible if nearly all of the internal degrees of freedom become locked to each other by the nonlinear interactions following the “slaving principle” developed by H. Haken in the 1960s and 1970s \cite{83,84}. The particular pattern itself is selected by choosing suitable values for one or at most a few parameters. Because the slaving principle eliminates the need to control a vast number of degrees of freedom, the number of required processing steps is reduced to adjusting a few control parameters. This, in turn, reduces the time needed to print a structure in case of a printer exploiting self-organisation. In a sense, self-organisation automatically achieves a vast degree of parallelisation, as the internal degrees of freedom organise themselves through their nonlinear interactions \cite{61,85}.

The locking of the internal degrees of freedom is best illustrated by the well-studied example of locking of the frequency modes in a laser during the onset of modelocking \cite{73}. A laser’s cavity imposes discrete longitudinal modes, equally spaced in frequency. In an ordinary multimode laser, a number of these modes are excited and oscillate independently of each other with random relative phases. In the case of modelocking, thousands to millions of modes spontaneously lock-up, \emph{i.e.}, they develop fixed phase relationships. This corresponds to forming a coherent structure in time, namely, an ultrashort pulse. Various interrelated formulations have been developed, whereby differential equations are typically used by taking the continuum limit of the vast number of modes. To better illustrate the slaving principle, here, we retain the discrete modes \cite{86}. The electric field within the cavity can be written as
\begin{equation}
E(z,t)=\sum_m a_m (t)e^{-i(\omega_0+\omega_R m)t} e^{(k_0 + \Delta k m)z} +c.c.,
\end{equation}
where $\Delta k = 2\pi/L$, $\Omega=\nu_g \Delta k$, and $k_0$ and $\omega_0$ are the central wavenumber and frequency, respectively.  Then, the equations of evolution for these modes can be written as
\begin{equation}
\dot{a}_m=(i\gamma_d - \gamma_g)m^2 a_m + (i\gamma_k - \gamma_s) \sum_{j-k+l=m}^N a_j a_k^* a_l + g a_m
\end{equation}
where the parameters, $\gamma_d$, $\gamma_g$, $\gamma_k$, $\gamma_s$, and $g$ correspond to various effects, such as gain, nonlinearities and  $N$ is the maximum number of frequency modes the laser cavity supports. Modelocking is obtained for certain values of these parameters. A special case with an analytic solution is the formation of a soliton-like pulse \cite{Haus_soliton}. All the $2N$ degrees of freedom corresponding to the amplitude and phase of $a_m$ become fixed, save for noise due to quantum fluctuations or of technical origin. In the time domain, a pulse emerges with the form, $A_p {\rm sech}(t/\tau_p)$, where $A_p$ and $\tau_p$ are its amplitude and temporal width, respectively, as determined {\rm collectively} by $a_m$. In this case, the so-called soliton area theorem provides a straightforward relation that we can conveniently use to illustrate the principle of slaving. The amplitude and temporal width of the soliton are interrelated by $A_p^2 \tau = \gamma_d/\gamma_k$, and the energy of the soliton is $E_p=A_p^2\tau_p$, where, in both cases, we neglected proportionality constants of the order of $1$. Let’s take a moment to contemplate this result. A modelocked laser contains very many modes or degrees of freedom. After the onset of modelocking, a coherent structure arises. Its emergent characteristics, such as width and amplitude, are determined collectively by these many degrees of freedom. A handful of parameters influence these characteristics; for instance, by controlling the pump power of the laser, which sets $E_p$ and by changing the dispersion of the cavity, $\gamma_d$, we can independently set the amplitude and width. This is how up to millions of degrees of freedom, $a_m$, become fixed up to an overall phase. Still, many of the overall macroscopic characteristics of the pulse can be controlled through the two parameters of pump power and dispersion alone.

\subsection{Taming the \emph{Mischief of Fluctuations}}

The importance of fluctuations seems to have been appreciated only when it was desired to suppress them. The notion that fluctuations can be an excellent control tool appears to go unnoticed. This is hardly surprising given that the principal aim of the large engineering field of control theory is to suppress noise. While this is undoubtedly desirable for linear and near-equilibrium systems, as we will argue below, this is not always the case for self-organisation, where noise, fluctuations, defects, and all other incoherent modulations (or even coherent ones when they are undesired or unexpected) can increase robustness, improve evolutionary fitness, enhance functionality, or is outright necessary for emergence.

When a spatiotemporal energy flux is introduced to a system at thermal equilibrium, first, dissipation breaks detailed balance, leading to a net flow (of mater, chemical reactions, trajectories of particles, probabilities, {\em etc}.) both in space and in time. This leads to significant amplification of the fluctuations, often called critical fluctuations or giant number fluctuations. Unlike in closed systems (already at or approaching thermal equilibrium), the fluctuations become correlated in open systems (sustained far from equilibrium), ergo, do not satisfy the central limit theorem. As a result, not only is time-reversibility broken, but new forms of nonlinearities arise from the newly formed correlations. Under such conditions, the system is not stationary; it is driven from one state to another. The number of states that a complex system can be driven to is directly related to its nonlinear interactions with the net flux. High nonlinearity translates into an expanding phase space with an abundant number of states. However, opening up the phase space to exploration by the system does not guarantee that the system will be able to visit a wide variety of states. For that, it needs fluctuations. Fluctuations have to be strong enough to “shake” the phase space to turn the static steady-states into dynamic ones so that there is no single thermodynamic minimum energy state but a multitude of them, constantly evolving. Undisputedly, there is no static energy landscape anymore; instead, borrowing the concept from evolutionary biology, there is a “fitness landscape.” 

A useful conceptual tool for understanding the evolution of a self-organising system comes from evolutionary biology, namely, the notion of a fitness landscape \cite{87,88}, first introduced by John Maynard Smith in 1970. A fitness landscape can be thought of as a generalisation of the potential surfaces we have for conservative mechanical systems. In evolutionary biology, fitness represents the net reproductive rate, determining the success of a particular genomic sequence in a population. The coordinates of this landscape correspond to the different configurations of the genomic sequence. Unlike the mechanical analogy, the landscape is not static but changes dynamically as the system evolves. A population, an ensemble of many individual organisms with different genomic sequences, tend to evolve towards higher fitness values. In the analogy to an evolving self-organising system, the landscape coordinates correspond to different values in the system’s phase space. In place of fitness, we have the growth rate of a given pattern. In both cases, the system tries to climb the peaks in their high-dimensional landscapes. 

\begin{figure}[t]
\includegraphics[width=\textwidth]{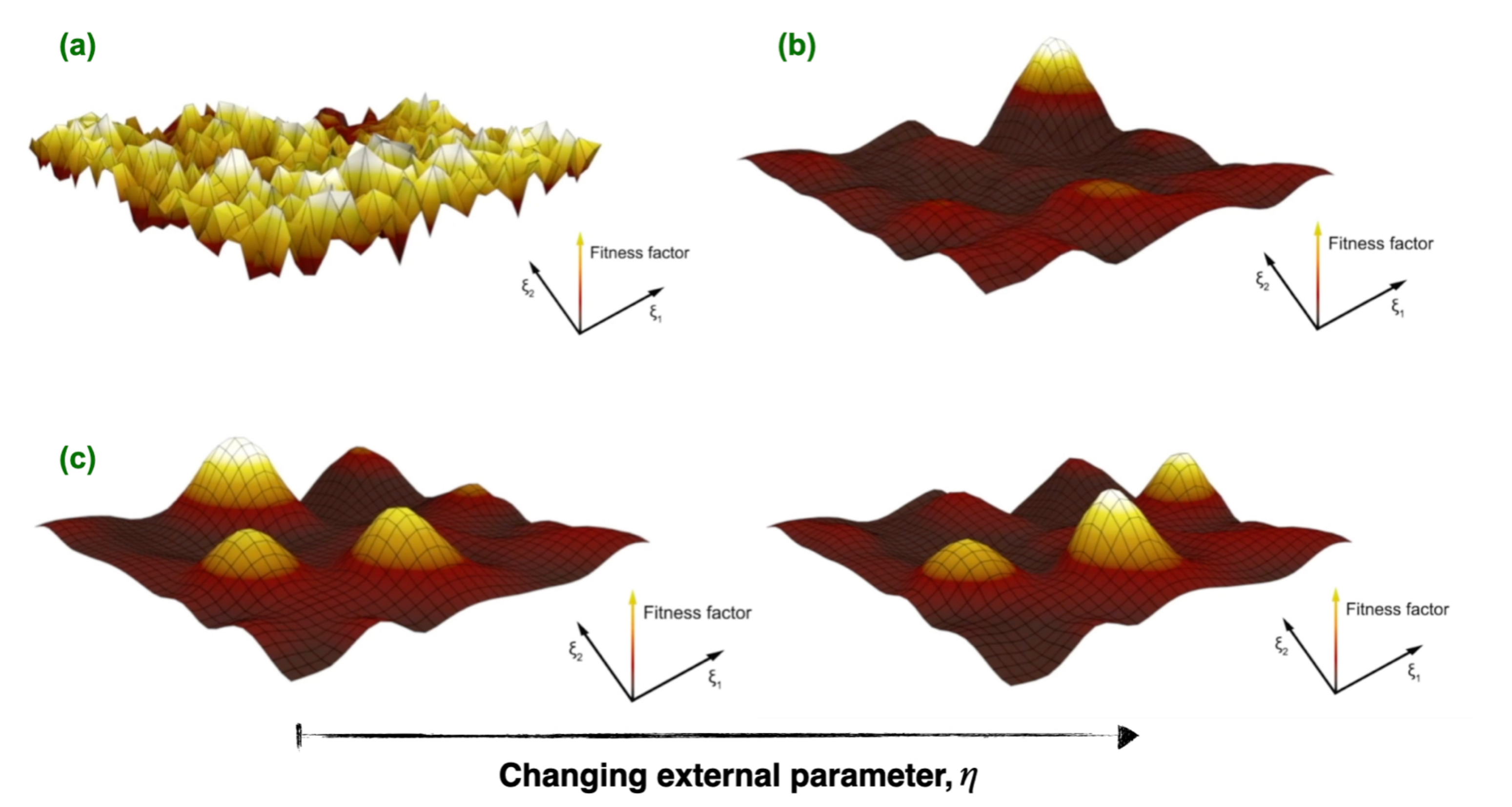}
\caption{(a) A Badlands landscape has many peaks of comparable size, whereas (b) a Fujiyama fitness landscape has a dominating peak. In the context of a 3D atom printer, an intermediate landscape with several peaks corresponding to a different atomic structure is of greater interest. The most prominent  peak must have a sufficient margin. The fluctuations must be present and strong enough to ensure that the system reliably converges to the most prominent peak each time it is started from different initial coordinates. (c) By changing the control parameters, such as the power of the laser beam that drives the self-organised process, the landscape is deformed to make the peak corresponding to the desired atomic structure as the most prominent peak.}
\label{fig:landscapes}     
\end{figure}

These fitness landscapes may have many peaks of comparable size, known as a Badlands landscape (fig. \ref{fig:landscapes}a), or it can have a single peak, known as a Fujiyama landscape (fig. \ref{fig:landscapes}b). Without fluctuations, a self-organised system will get stuck to whichever peak it happens to start close to, where each peak corresponds to a different self-organised pattern. The initial condition for a self-organised system corresponds to different starting coordinates in this landscape. Unless there is a single peak, as in the Fujiyama landscape, different initial conditions will have different peaks or patterns. Given our interest in controlling self-organisation to achieve a pattern, \emph{i.e.}, an atomic structure of our choosing, it is imperative that the parameters must be chosen such that there are multiple peaks in the landscape. Control of self-organisation can be understood in this analogy as changing the control parameters, such as the laser power, thus deforming the landscape to make a different peak, the tallest peak (fig. \ref{fig:landscapes}c). However, it is not guaranteed the system will find its way to the tallest peak, where there are multiple peaks. A rougher landscape is much more complex for the self-organising system to climb as the smaller local peaks form traps that the system cannot escape. In practice, this means that a different self-organised pattern, \emph{i.e.}, atomic structure, from what was intended may be erroneously obtained due to slight variations in the initial conditions. This is highly undesirable for the 3D atom printer.

Fluctuations come to our aid here. Any perturbation due to the fluctuations can be thought of as a little kick in the phase space, which can dislodge the system from being trapped at a small peak, allowing the self-organising system to explore its landscape more fully and evolve to the tallest peak. In this case, quite contrary to the case of direct writing, the presence of fluctuations is not detrimental but essential for the self-organisation to develop near independence from the initial conditions and to exhibit universality. Furthermore, since the fluctuations are part of the intrinsic feedback loops in complex natural systems and our experimental designs (fig. \ref{fig:size_no_matter} and \ref{fig:NLL_DSA_nano}), there is no need to suppress or control them extensively.

\section{From Self-organisation to a 3D Atom Printer}
\label{sec:4}

In this final section, we discuss how a 3D atom printer may be constructed with the aid of self-organisation. As mentioned in Section \ref{sec:2}, 3D printer technology based on direct writing is highly developed and versatile at the microscopic scale but can barely reach the mesoscopic scale. Three hard limitations arising from fundamental physical effects prevent further scaling down the resolution. At the same time, as discussed in Section \ref{sec:3}, laser-driven self-organisation is not directly limited by the same three challenges. As a result, self-organisation allows the creation of materials with control over the structure down to the atomic scale.

\begin{figure}[t]
\includegraphics[width=\textwidth]{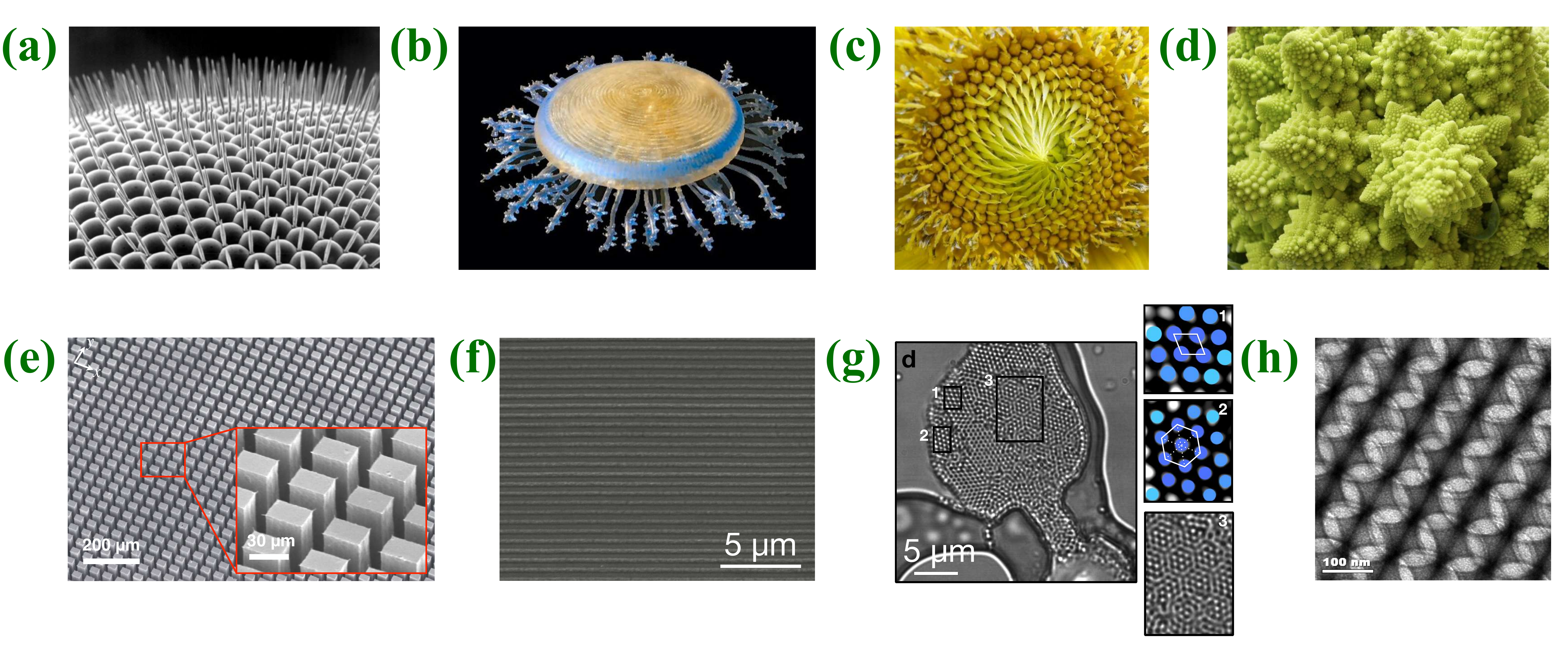}
\caption{Examples to self-organised structures exhibited by living organisms: (a) fly eye (Image Credit: Unknown), (b) Jellyfish Porpita mediterranea (Image Credit: Max Mackenzie), (c) sunflower (Image Credit: Unknown), (d) cauliflower (Image Credit: WindyPower, Deviant Art). Examples of self-organised structures we have created in the laboratory: (e) 3D structures created in silicon \cite{40}, (f) 2D patterns created on titanium \cite{32}, (g) self-assembled colloids \cite{57}, (h) a self-assembled hierarchical zeolite structure.}
\label{fig:nature_ildaylab}     
\end{figure}

In Nature, highly complex structures spontaneously arise through self-organisation (fig. \ref{fig:patterns}). This is observed ubiquitously at all scales. The most elaborate structures ever observed are, arguably, those of biological organisms and, among them, probably, the brain. Biological structures grow through what may be thought of as a myriad network of multiple self-organised processes, often arranged hierarchically and developed by evolution over aeons. The relatively simpler inorganic but highly complex structures and living organisms prove that complex structures can be created purely by self-organisation. We believe that a 3D printer operating similarly is possible, where the printing processes do not involve point-by-point processing at any step or scale, quite unlike existing 3D printers. Instead, the intended object is printed entirely by self-organised processes, which are still externally driven. The conditions and the parameters that drive vary over time, but with the complete elimination of the point-by-point process, the final resolution of the structure is not explicitly determined at all. However, our present understanding of self-organised processes is in its infancy. We have barely begun to understand how to steer a single self-organised process to achieve a specific pattern. Complex self-organisation at multiple scales requires hierarchical self-organisation, which is even less understood. The perspective for a 3D atom printer that we discuss here is that of a simpler but better-understood arrangement. In this case, self-organisation has the more modest role of bridging the mesoscopic scale, where a direct-writing approach is permissible, and the nanometric and atomic scales (fig. \ref{fig:nature_ildaylab}). 

\begin{figure}[t]
\includegraphics[width=\textwidth]{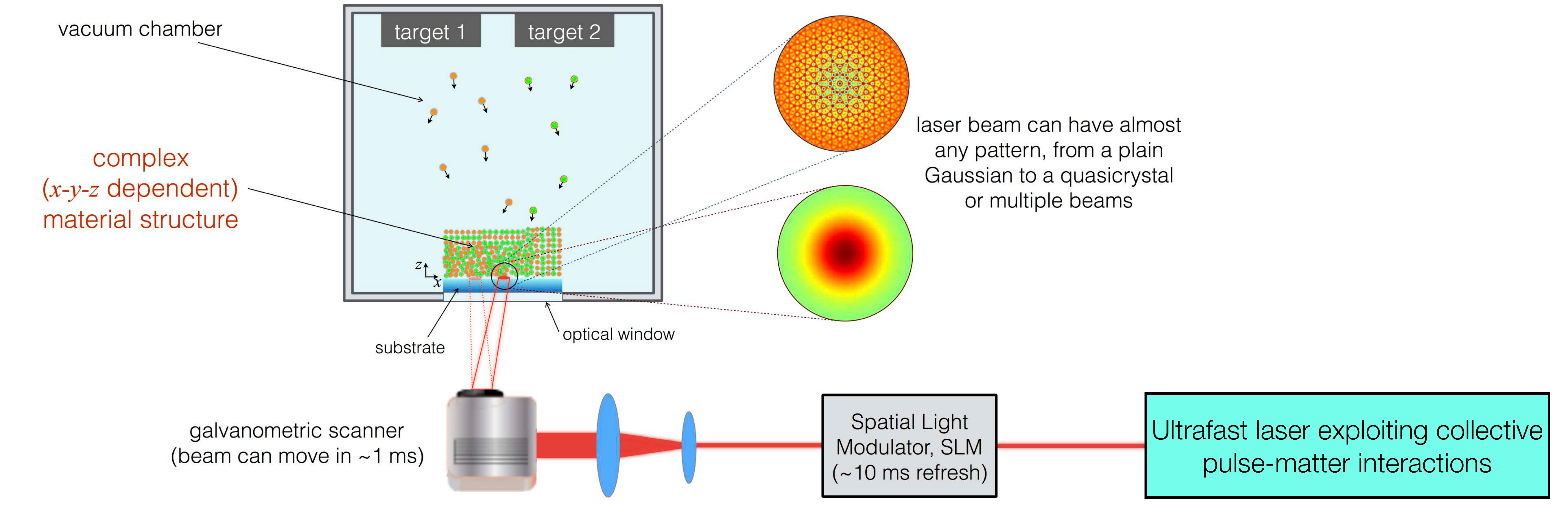}
\caption{A possible implementation of a 3D atom printer using vacuum deposition. Multiple material targets are used as sources of the different atoms and molecules for the intended composition. These targets could be activated by magnetrons or by a pulsed laser. The 3D object is printed on a substrate layer by layer. The atomic structure and composition are controlled by spatiotemporal thermal gradients formed on the substrate using dynamically controlled structured light from an ultrafast laser.}
\label{fig:atom_printer}     
\end{figure}

In order to make our discussion more concrete, we next discuss the broad outlines of a specific implementation of a 3D atom printer that could be built using existing technology. We consider a setup (fig. \ref{fig:atom_printer}) that is broadly based on our earlier demonstration of vacuum deposition with a static spatial thermal gradient \cite{80} but introduces the dynamic spatiotemporal thermal gradients of the dissipative dynamic self-assembly \cite{57,61}. The elements making up the material to be deposited is generated by ablation from targets inside the chamber. There may be multiple targets, but we depict two here for simplicity. These targets may be, for instance, Si and SiO$_2$ \cite{80}. Ablation of atoms from these targets may be by magnetrons or pulsed lasers for greater flexibility. In the latter case, as many targets can be addressed as needed by an additional arm of the same laser that drives the self-organisation process. This way, the ablation starting and stopping times of each target can be precisely controlled synchronously with the beam pattern applied to the object’s surface being printed.   

The spatial thermal gradients on the surface of the object will be determined by the laser beam, which can be structured to create virtually any pattern. The diffraction limit only limits the most miniature scale of these patterns, but as we have discussed in the previous section, this does not preclude control over the atomic structure. The main imprint of diffraction will be to limit the minimum distance between two regions of different intended atomic structures. This is nearly the same limit as the voxel size, \emph{i.e.}, the smallest independently addressable region. 

During any layer-by-layer deposition process, the growth direction is mapped to time, as in the common forms of 3D printing and vacuum deposition. This mapping readily creates an additional opportunity for bypassing the optical diffraction limit. The speed at which the driving laser beam is reconfigured compared to the deposition rate sets a lower limit to the spatial resolution along the vertical direction. Currently, spatial light modulators (SLMs) offer the highest number of independent actuators, but they are relatively slow with refresh rates of $\sim\!\!10$ ms. Deformable mirror and MEMS technologies provide response times of $\sim\!\!10 \mu$s. They have relatively fewer (several thousand instead of several million), but still a large number of actuators. Even at the present speed limits, hardly an atomic layer is deposited within the time scale for being able to reconfigure the laser beam. Furthermore, it is reasonable to expect significant advances in these parameters. Therefore, the printing resolution is not limited by diffraction or laser physics, rather by the dynamics of the deposition and following self-organisation processes. In theory, control down to the single atomic layer is plausible under the right conditions. 

Next, we discuss how the {\em Explosion of Complexity} is circumvented in this implementation, which paves the way to reasonable printing times for macroscopic objects. We aim to provide an order of magnitude calculation to illustrate that reasonable printing times are feasible. Consider, again, the same vacuum deposition setup of fig. \ref{fig:atom_printer}. Even disregarding likely progress in the deformable mirror or SLM technologies, the former allows to reconfigure the spatial thermal gradients on the surface within $\sim\!\!10 \mu$s. If the deposition rate is adjusted to one atom layer on average during this time, printing rates of $\sim\!\!0.1$ mm/s are already conceivable while updating the structured laser field once per printing of each atomic layer. Thus, a 1-mm-thick structure could be printed over a few minutes. The lateral size of the object does not impact the processing time as long as it is not wider than the field of view covered by the structured laser beam. Therefore, 3D structures approaching a 1 mm$^3$ in size could conceivably be printed in the timescale of minutes. As discussed in Section \ref{sec:3}, the expected time for an equivalent-sized object is many orders of magnitude longer if attempting to process point by point in atomic-scale resolution, even if it were possible.

Finally, the impact of fluctuations is entirely different in the case of self-organisation. When no attempt is made to control the exact position and state of the individual particles, the fluctuations play an integral and welcome role in many self-organised or other feedback-driven and iterative processes. There is a priori no reason to avoid or suppress them, and there are many cases where the presence and characteristics of the fluctuations can alter the dynamically preferred atomic arrangement. Access to atomic structures that are not reachable through near-equilibrium states can enable new functionalities, for example, through noise-induced transitions \cite{89}.

Overall, we believe there is a possible path towards a 3D atom printer by fully exploiting self-organisation. The entirely self-organised creation of a complex 3D object, much as the organic growth of a biological organisation, except that the details are externally steered rather than set by its evolutionary history, is a distant possibility. We believe that our present understanding of the fundamentals of self-organisation is too primitive even to conjecture on this possibility beyond expressing our belief that it should be possible. However, as discussed above, exploiting self-organisation to bridge the gap between the mesoscopic scale, where direct writing is possible, and the atomic scale, appears to us a perfectly viable path. We further believe that the first demonstration of such a device could be achieved within a decade. To this end, we need a more advanced understanding of the physics of laser-driven self-organisation. We will also need better experimental tools. These tools include interaction chambers, which may not be limited to traditional vacuum deposition techniques but may include, for example, microplasmas for the generative, highly reactive species. However, the most important role will likely be played by ultrafast lasers designed specifically for driving self-organised processes in the matter they interact with, possibly by exploiting collective pulse-matter interactions tailored to achieve a degree of separate control over the electron and atom (lattice) temperatures. 

In concluding, we look forward to an exciting decade of a creative combination of curiosity-driven fundamental advances in self-organisation physics with advanced laser technologies and innovative developments in 3D printing and material deposition techniques with disruptive and immediate industrial applications.  

\begin{acknowledgement}
The authors acknowledge funding the European Research Council (ERC) through the Ph.D. (grant no. 853387) NLL (grant no. 617521), and the SUPERSONIC (grant no. 966846) projects and the T\"{U}B\.{I}TAK – The Scientific and Technological Research Council of Turkey through projects no. 20AG024, and 20AG001.
\end{acknowledgement}
%

%%%%%%%%%%%%%%%%%%%%%%%% referenc.tex %%%%%%%%%%%%%%%%%%%%%%%%%%%%%%
% sample references
% %
% Use this file as a template for your own input.
%
%%%%%%%%%%%%%%%%%%%%%%%% Springer-Verlag %%%%%%%%%%%%%%%%%%%%%%%%%%
%
% BibTeX users please use
% \bibliographystyle{}
% \bibliography{}
%

\end{document}